\documentclass[amsmath,amssymb,superscriptaddress,nobalancelastpage,aps,prb,twocolumn]{revtex4-1}

\usepackage{graphicx}
\usepackage{varioref}
\usepackage{xr-hyper}
\usepackage{xcolor}
\usepackage{nicefrac}
\usepackage{xfrac}
\usepackage{hyperref}
\hypersetup{colorlinks,linkcolor=blue,urlcolor=blue,citecolor=blue}
\usepackage{ulem}

\begin{document}

\title{Damped spin excitations in a doped cuprate superconductor with orbital 
hybridization}
      
\author{O.~Ivashko}
\affiliation{Physik-Institut, Universit\"{a}t Z\"{u}rich, Winterthurerstrasse 
190, CH-8057 Z\"{u}rich, Switzerland}

\author{N.~E.~Shaik}
\affiliation{Institute for Condensed Matter Physics, \'{E}cole Polytechnique 
Fed\'{e}rale de Lausanne (EPFL), CH-1015 Lausanne, Switzerland}
	
\author{X.~Lu}
\affiliation{Swiss Light Source, Paul Scherrer Institut, CH-5232 Villigen PSI, 
Switzerland} 

\author{C.~G.~Fatuzzo}
\affiliation{Institute for Condensed Matter Physics, \'{E}cole Polytechnique 
Fed\'{e}rale de Lausanne (EPFL), CH-1015 Lausanne, Switzerland}

\author{M.~Dantz}
\affiliation{Swiss Light Source, Paul Scherrer Institut, CH-5232 Villigen PSI, 
Switzerland}

\author{P.~G.~Freeman}
\affiliation{Jeremiah Horrocks Institute for Mathematics, Physics and 
Astronomy, University of Central Lancashire, PR1 2HE Preston, United Kingdom}

\author{D.~E.~McNally}
\affiliation{Swiss Light Source, Paul Scherrer Institut, CH-5232 Villigen PSI, 
Switzerland}
 
\author{D.~Destraz}
\affiliation{Physik-Institut, Universit\"{a}t Z\"{u}rich, Winterthurerstrasse 
190, CH-8057 Z\"{u}rich, Switzerland}
 
\author{N.~B.~Christensen}
\affiliation{Department of Physics, Technical University of Denmark, DK-2800 
Kongens Lyngby, Denmark.}
 
\author{T.~Kurosawa}
\affiliation{Department of Physics, Hokkaido University - Sapporo 060-0810, 
Japan}
 
\author{N.~Momono}
\affiliation{Department of Physics, Hokkaido University - Sapporo 060-0810, 
Japan}
\affiliation{Department of Applied Sciences, Muroran Institute of Technology, 
Muroran 050-8585, Japan}

\author{M.~Oda}
\affiliation{Department of Physics, Hokkaido University - Sapporo 060-0810, 
Japan}
 
\author{C.~E.~Matt}
\affiliation{Physik-Institut, Universit\"{a}t Z\"{u}rich, Winterthurerstrasse 
190, CH-8057 Z\"{u}rich, Switzerland}
 
\author{C.~Monney}
\affiliation{Physik-Institut, Universit\"{a}t Z\"{u}rich, Winterthurerstrasse 
190, CH-8057 Z\"{u}rich, Switzerland}

\author{H.~M.~R{\o}nnow}
\affiliation{Institute for Condensed Matter Physics, \'{E}cole Polytechnique 
Fed\'{e}rale de Lausanne (EPFL), CH-1015 Lausanne, Switzerland}

\author{T.~Schmitt}
\affiliation{Swiss Light Source, Paul Scherrer Institut, CH-5232 Villigen PSI, 
Switzerland}

\author{J.~Chang}
\email{johan.chang@physik.uzh.ch}
\affiliation{Physik-Institut, Universit\"{a}t Z\"{u}rich, Winterthurerstrasse 
190, CH-8057 Z\"{u}rich, Switzerland}

\begin{abstract}
A resonant inelastic x-ray scattering study of  overdamped 
spin excitations in slightly underdoped La$_{2-x}$Sr$_{x}$CuO$_4$ (LSCO)  with 
$x=0.12$ and $0.145$ is presented. Three high-symmetry directions have been 
investigated: (1) the antinodal $(0,0)\rightarrow (\nicefrac{1}{2},0)$, (2) the 
nodal $(0,0)\rightarrow (\nicefrac{1}{4},\nicefrac{1}{4})$ and (3) the zone 
boundary direction $(\nicefrac{1}{2},0)\rightarrow 
(\nicefrac{1}{4},\nicefrac{1}{4})$ connecting  these two. The overdamped 
excitations exhibit strong dispersions along (1) and (3), whereas a much more 
modest dispersion is found along (2). This is in strong contrast to the undoped 
compound La$_{2}$CuO$_4$ (LCO) for which the strongest dispersions are found 
along (1) and (2). The  $t-t^{\prime}-t^{\prime\prime}-U$ Hubbard model used to 
explain the excitation spectrum of LCO predicts -- for constant $U/t$ -- that 
the dispersion along (3) scales with $(t^{\prime}/t)^2$. However, the diagonal 
hopping $t^{\prime}$ extracted on LSCO using single-band models is low 
($t^{\prime}/t\sim-0.16$) and decreasing with doping. We therefore invoked a 
two-orbital ($d_{x^2-y^2}$ and $d_{z^2}$) model which implies that $t^{\prime}$ 
is enhanced. This effect acts to enhance the zone-boundary dispersion within the Hubbard model. We 
thus conclude that hybridization of $d_{x^2-y^2}$ and $d_{z^2}$ states has a 
significant impact on the zone-boundary dispersion in LSCO.
\end{abstract}

\maketitle

\section{Introduction} 
Considerable research is being undertaken in the quest to reach consensus on the 
mechanism of high-temperature superconductivity~\cite{LeeRMP06} and the 
associated pseudogap phase~\cite{NormanAP05} in copper-oxide materials 
(cuprates). The energy scales governing the physical properties of these layered 
materials therefore remain of great interest. It is known that these materials 
are characterized by a strong superexchange interaction $J_1=4t^2/U$ where $t$ 
is the nearest-neighbor hopping integral and $U$ is the Coulomb interaction. To 
first order, this energy scale sets the bandwidth of the spin-excitation 
spectrum. Resonant inelastic x-ray scattering (RIXS) 
experiments~\cite{AmentRMP11} have demonstrated that this bandwidth stays 
roughly unchanged across the entire phase diagram~\cite{TaconNATP11,DeanNATM13} 
of hole doped cuprates. It has also been demonstrated that the cuprates belong 
to a regime (of $t$ and $U$) where the second-order exchange interaction 
$J_2=4t^4/U^3$ contributes to a  spin-excitation dispersion along the 
antiferromagnetic zone boundary 
(AFZB)~\cite{ColdeaPRL01,HeadingsPRL10,DelannoyPRB2009,DallaPiazzaPRB12}. 
Moreover, it is known from band structure calculations and experiments that the 
next nearest-neighbor (diagonal) hopping integral $t^{\prime}$ constitutes a 
non-negligible fraction of $t$~\cite{YoshidaPRB06}. 
Empirically~\cite{PavariniPRL01}, the superconducting transition scales with the 
ratio $t^{\prime}/t$ whereas Hubbard type models predict the opposite 
trend~\cite{WhitePRB99,MaierPRL00}. As a resolution, a two-orbital model -- in 
which hybridization of $d_{z^2}$ and $d_{x^2-y^2}$ states suppresses $T_c$ and 
enhances $t^{\prime}$  -- has been put forward~\cite{SakakibaraPRL10}.

Here, we address the question as to how $t^{\prime}$ influences the 
spin-excitation spectrum at, and in the vicinity of, the antiferromagnetic zone 
boundary. We have therefore studied -- using the RIXS technique -- slightly 
underdoped compounds of  La$_{2-x}$Sr$_x$CuO$_4$ (LSCO) with $x=0.12$ and 
$0.145$. Even though the system is not antiferromagnetically ordered at these 
dopings, we  quantify the zone-boundary dispersion $\omega(\overline{q})$ by 
$E_{ZB}=\omega(\nicefrac{1}{2},0)-\omega(\nicefrac{1}{4},\nicefrac{1}{4})$. In 
doped LSCO a strongly enhanced zone-boundary dispersion is observed. As will 
also be shown, within the $t-t^{\prime}-t^{\prime\prime}-U$ Hubbard model, one 
generally expects that the zone-boundary dispersion scales with $t^{\prime}/t$ 
with a prefactor that depends on $U/t$. The Fermi-surface topology of LSCO, 
obtained from photoemission spectroscopy and analyzed with a single-band tight 
binding model, suggests that $t^{\prime}$ decreases with increasing 
doping~\cite{YoshidaPRB06,ChangPRB08a}. The Hubbard model is thus within a single-band 
picture not consistent with the experiment. However, using a two-orbital model, 
hybridization between $d_{z^2}$ and $d_{x^2-y^2}$ states enhances 
$t^{\prime}$~\cite{SakakibaraPRL10}. This provides a satisfactory description 
of the zone-boundary dispersion. We thus conclude that the two-orbital 
model~\cite{SakakibaraPRL10} is necessary to understand the spin-excitation 
spectrum of doped LSCO.

\begin{figure*}
	\begin{center}
		\includegraphics[width=0.995\textwidth]{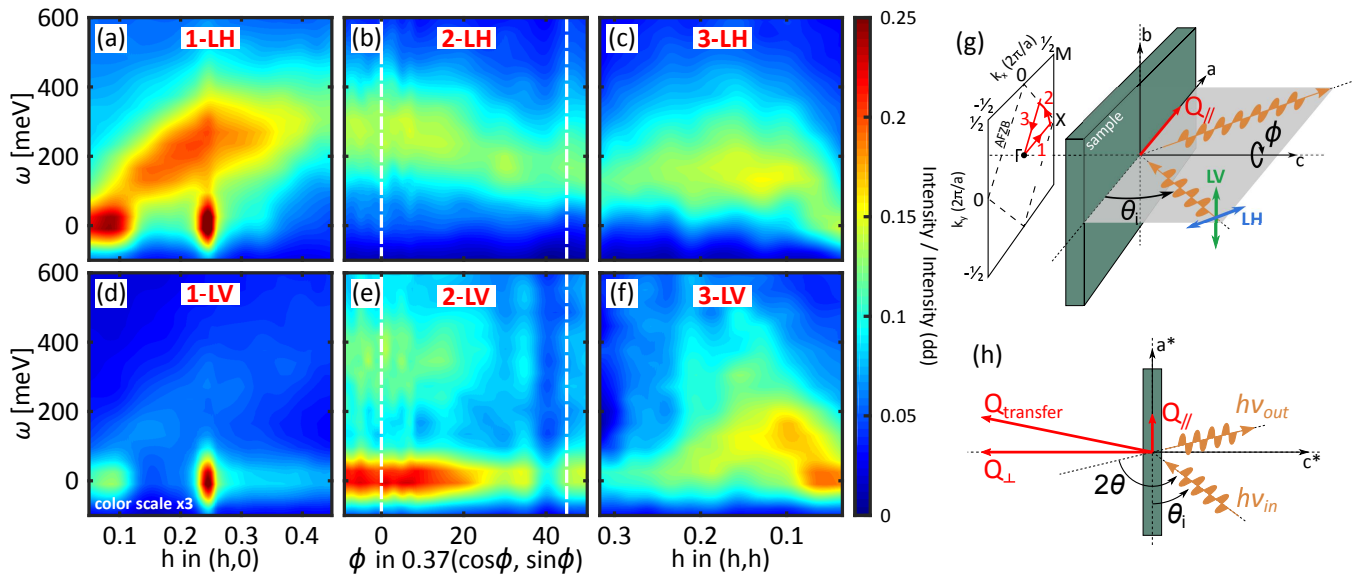}
	\end{center}
	\caption{(Color online) RIXS spectra versus momentum recorded on 
La$_{2-x}$Sr$_x$CuO$_4$ with  $x=0.12$ under grazing exit conditions and 
displayed in false color scale for different light polarizations. (a,d) RIXS 
intensity maps along the antinodal direction for linear-horizontal and 
linear-vertical incident light polarizations.  (c,f) similar maps but along the 
nodal direction. (b,e) azimuthal  RIXS maps connecting nodal and antinodal 
directions as shown schematically in (g). Consistent with what has previously 
been shown the spin-excitation matrix element is strongest for the LH 
polarization. By contrast, the charge-density-wave reflection at 
$Q_{\mathrm{CDW}}=(\pm\delta_1,\delta_2)$ with $\delta_1\sim0.25$ and 
$\delta_2\sim0.01$ is about three times more intense with LV polarization. Panels (g) and 
(h) display the scattering geometry (side and top view respectively) where 
$\theta_{i}$ indicates the incident angle and $\phi$ is the azimuthal angle. 
Varying these angles allows us to scan the in-plane momentum 
$Q_{\!\textfractionsolidus\!\textfractionsolidus}$. In (g) scan directions, with 
respect to the antiferromagnetic zone boundary are shown.} \label{fig:fig1}
\end{figure*}

\begin{figure*}
 	\begin{center}
 		\includegraphics[width=0.995\textwidth]{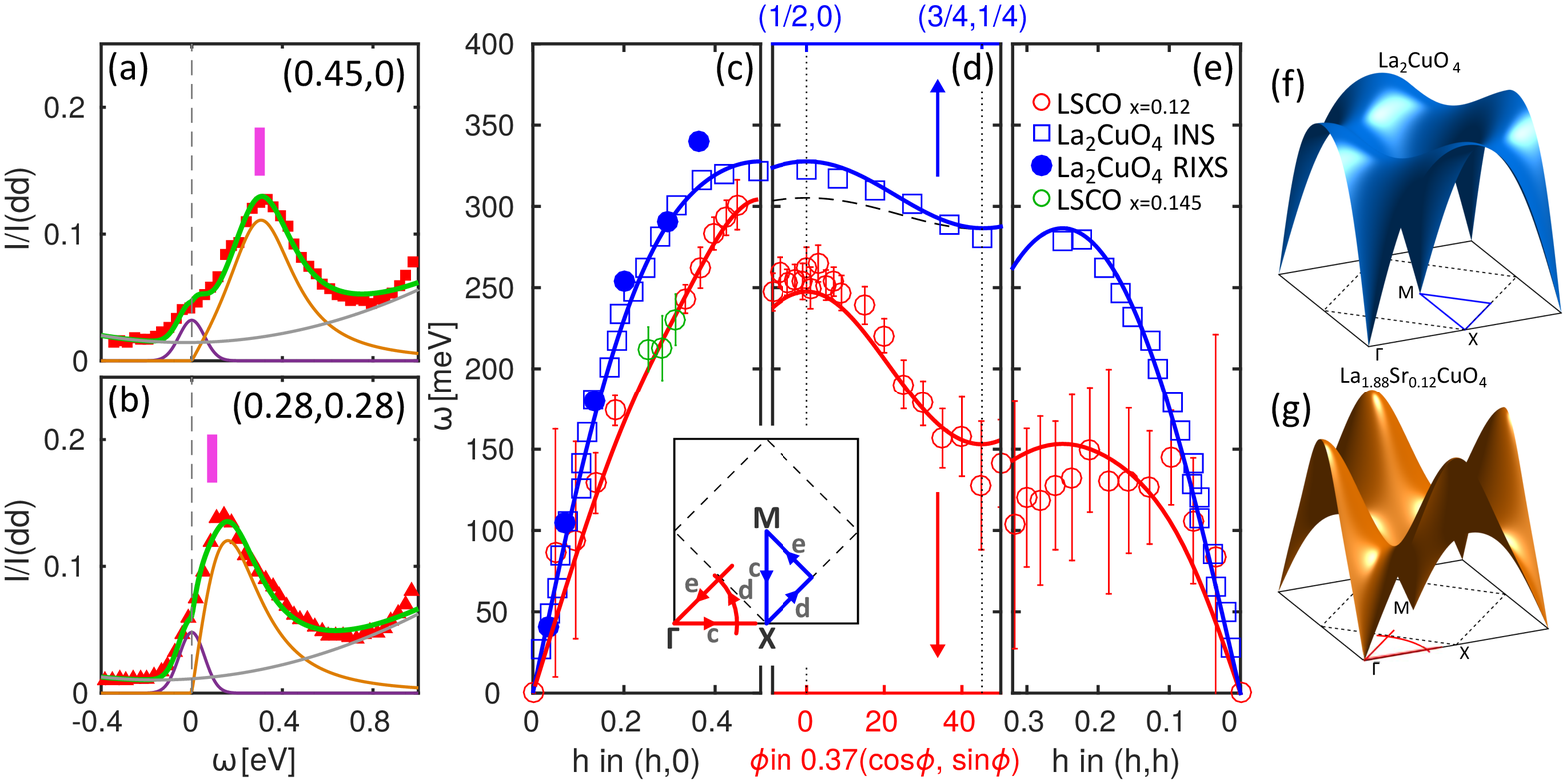}
 	\end{center}
 	\caption{(Color online) RIXS spectra for antinodal (a) and nodal (b) 
directions with the indicated in-plane momentum. The fit (solid green curve) is 
composed of three components: elastic line (purple), spin excitation (orange) 
modeled by an antisymmetric Lorentzian function  and a quadratic background 
(grey) -- see text for more detailed information. Vertical bars indicate the 
obtained poles of the Lorentzian function. (c)-(e) dispersion of the magnetic 
excitations in La$_2$CuO$_4$  observed by neutron scattering (open blue squares 
-- Ref.~\onlinecite{HeadingsPRL10}) and RIXS (filled blue circles -- 
Ref.~\onlinecite{BraicovichPRL10}) and La$_{2-x}$Sr$_x$CuO$_4$  with $x=0.12$ 
(red circles) measured by RIXS (this work). Green circles in (c) are extracted 
from La$_{2-x}$Sr$_x$CuO$_4$  with $x=0.145$ data. Within the antiferromagnetic 
zone scheme (indicated by the dashed line in the insert), red and blue cuts $c$ 
and $e$ are the equivalent antinodal and nodal directions. Solid lines in 
(c)-(e) are fits using a Heisenberg model, see text for further explanation. In 
(d) thin dashed line is the correspective azimuthal scan, for La$_2$CuO$_4$, 
extracted from the above mentioned model. (f)-(g) schematic illustration of the 
spin-excitation dispersion in La$_{2-x}$Sr$_{x}$CuO$_4$ with $x=0$ and $x=0.12$, 
as indicated. In the doped compound, the spin-excitation dispersion is strongly 
renormalized along the diagonal (nodal, $\Gamma$-M) direction. Blue and red 
patterns indicate the experimentally measured high-symmetry directions.}
 	\label{fig:fig2}
\end{figure*}

\section{Method}
The RIXS experiment was carried out at the ADvanced RESonant Spectroscopies 
(ADRESS) beamline~\cite{ghiringhelliREVSCIINS2006,strocovJSYNRAD2010} at the 
Swiss Light Source (SLS) with the geometry shown in Fig.~\ref{fig:fig1}(h). The 
newly installed CARVING RIXS manipulator allowed us to probe the full kinematically 
accessible reciprocal space $\overline{q}=(h,k)$ with a scattering angle of $130^\circ$. 
Incident photons with an energy of $933$~eV (at the Cu $L_{3}$-edge resonance) 
gave an instrumental energy and momentum resolution of $132$~meV and $0.01$ 
\AA$^{-1}$ respectively. Both the linear horizontal (LH) and linear vertical 
(LV) light polarizations were applied to probe high quality single crystals  of 
La$_{2-x}$Sr$_{x}$CuO$_4$  with $x=0.12$ and $0.145$ ($T_c=27$ and $35$~K 
respectively). These crystals were grown by the traveling floating zone 
method~\cite{KomiyaPRB02} and previously characterized in 
neutron~\cite{ChangPRL09,RomerPRB13,ChangPRB12} and muon spin-resonance ($\mu$SR)~\cite{ChangPRB08} 
experiments. Ex situ prealignment of the samples was carried out using a Laue 
diffractometer. The samples were cleaved in situ using a standard top-post 
technique and all data were recorded at $T=20$~K. Although being in the low 
temperature orthorhombic (LTO) crystal structure, tetragonal notation 
$a{\cong}b\approx3.78$ \AA ($c\approx13.2$ \AA) is adopted to describe the 
in-plane momentum $(h,k)$ in reciprocal lattice units $2\pi/a$.

\section{Results}
Fig.~\ref{fig:fig1}(a-c) displays grazing exit RIXS spectra of 
La$_{1.88}$Sr$_{0.12}$CuO$_4$ recorded with incident LH light polarization along 
three trajectories as indicated in (g). Data along the same directions but 
measured with incident LV polarization are shown in (d-f). Besides the strong 
elastic scattering found at the specular condition $[Q=(0,0)]$, an elastic 
charge-density-wave (CDW) reflection is found -- consistent with existing 
literature~\cite{ThampyPRB2014,HaydenPRB2014} -- along the $(h,0)$ direction at 
$Q_{\mathrm{CDW}}=(\delta_1,\delta_2)$ with $\delta_1=0.24(6)$ and $\delta_2\simeq0.01$. The 
charge order reflection serves as a reference point, demonstrating precise 
alignment of the crystal.

For grazing exit geometry, it has previously been demonstrated that 
spin excitations are enhanced in the LH channel~\cite{TaconNATP11}. In 
Fig.~\ref{fig:fig2}(a,b), selected raw RIXS spectra recorded with LH 
polarization are shown for momenta near the $(\nicefrac{1}{2},0)$ and 
$(\nicefrac{1}{4},\nicefrac{1}{4})$ points. The low-energy part of the spectrum 
consists of three components: a weak  elastic contribution, a smoothly varying 
background and a damped spin excitation. It is immediately clear that the 
excitations near $(\nicefrac{1}{4},\nicefrac{1}{4})$ are significantly softened 
compared to those observed around the $(\nicefrac{1}{2},0)$-point (see 
Fig.~\ref{fig:fig2}(a,b)).

For a more quantitative analysis of the magnon dispersion, we modeled the 
elastic line with a Gaussian for which the standard deviation $\sigma=56$~meV 
was set by the instrumental energy resolution. A second order polynomial 
function is used to mimic the background. Finally, to analyze the 
spin excitations we adopted the response function of a damped harmonic 
oscillator~\cite{TaconNATP11,MonneyPRB16,LamsalPRB16}: 
\begin{align}
\chi^{\prime\prime}(\omega) & =\chi_0^{\prime\prime}  \frac{\gamma\omega 
}{\left( \omega^2-\omega_0^2 \right) ^2+\omega^2\gamma^2}\nonumber\\ &= 
\frac{\chi_0^{\prime\prime}}{2\omega_1}\left[\frac{\gamma/2}{ 
(\omega-\omega_1)^2+(\gamma/2)^2}-\frac{\gamma/2}{ 
(\omega+\omega_1)^2+(\gamma/2)^2}\right], \nonumber
\end{align}
where the damping coefficient $\gamma/2=\sqrt{\omega_0^2-\omega_1^2}$. The RIXS 
intensities are modeled by $\left[n_B(\omega)+1\right]
\chi^{\prime\prime}(\omega)$, where 
$n_B(\omega)=\left[\exp(\hbar\omega/k_BT)-1\right]^{-1}$ is the Bose factor. As 
shown in Fig.~\ref{fig:fig2}(a-b), fitting to this simple model provides a good 
description of the observed spectra. In this fashion, we extracted the 
spin-excitation pole dispersion $\omega_1(\overline{q})$ (Fig.~\ref{fig:fig2}(c-e)) along the 
three trajectories shown in the inset. To avoid the influence of CDW ordering 
on the spin-excitation dispersion~\cite{Miao17}, we analyzed around the charge 
ordering vector spectra of LSCO $x=0.145$ where charge order is absent. 

The extracted  spin-excitation dispersion of LSCO $x=0.12$ and $0.145$ is to be 
compared with the magnon dispersion of the parent compound 
La$_2$CuO$_4$~\cite{BraicovichPRL09,BraicovichPRL10,ColdeaPRL01,HeadingsPRL10}. 
Along the antinodal $(\nicefrac{1}{2},0)$ direction comparable dispersions are 
found. This is consistent with the weak doping dependence reported on 
LSCO~\cite{DeanNATM13} and the YBa$_2$Cu$_3$O$_{7-\delta}$ (YBCO) 
system~\cite{TaconNATP11}. For the nodal 
$(\nicefrac{1}{4},\nicefrac{1}{4})$ direction, the dispersion of the doped 
compound  is, however, strongly softened compared to La$_2$CuO$_4$. Whereas this 
effect has been reported for Bi-based~\cite{GuariseNATC2014,DeanPRB2014} and 
overdoped LSCO~\cite{MonneyPRB16}, we demonstrate directly by an azimuthal scan 
how exactly this softening appears. Notice that  the azimuthal dependence is 
closely related (but not exactly identical) to the scan along the 
antiferromagnetic zone boundary. 
 
\section{Discussion}
\begin{figure}
	\begin{center}
		\includegraphics[width=0.45\textwidth]{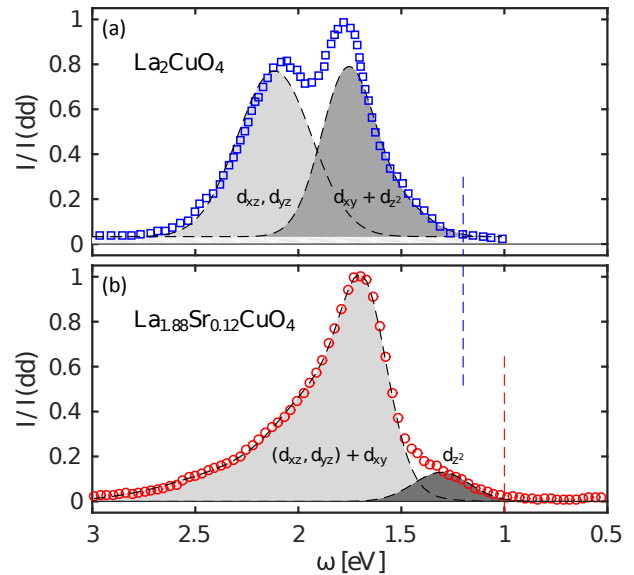}
	\end{center}
	\caption{(Color online) RIXS 
spectra showing the $dd$ excitations for La$_{2}$CuO$_4$ (a) (adopted from 
Ref.~\onlinecite{Peng16}) and La$_{1.88}$Sr$_{0.12}$CuO$_4$ (b) (this work). The 
grey shaded areas indicate schematically different orbital contributions. 
Vertical dashed lines display the onset of $dd$ excitations.}
	\label{fig:fig3}
\end{figure}

A recent systematic study~\cite{Peng16} of undoped cuprate compounds concluded 
that the zone-boundary dispersion scales with the crystal field splitting 
$E_{{z^2}}$ of the  $d_{x^2-y^2}$ and $d_{z^2}$ states. Exact numerical 
determination of  $E_{{z^2}}$ is still a matter of 
debate~\cite{SakakibaraPRL10,HozoiSREP11}.  For a tetragonal system,  
$E_{{z^2}}$  generally depends on the ratio between copper to apical and planar  
oxygen distances~\cite{MorettiNJP11}. The crystal field splitting 
$E_{z^2}$ can in principle be accessed by measuring the $dd$ excitations. For 
LCO, interpretations of the $dd$ excitations have consistently placed the 
$d_{z^2}$  level above (\textit{i.e.} closer to the Fermi level) both the  
$d_{xz,yz}$ and $d_{xy}$ states~\cite{MorettiNJP11,Peng16}. This is also 
consistent with density functional theory (DFT)~\cite{SakakibaraPRL10} and 
\textit{ab initio}~\cite{HozoiSREP11} calculations of the electronic band 
structure that find the $d_{z^2}$-band above the $t_{2g}$ states. In doped LSCO 
$x=0.12$, the spectral weight of the $dd$ excitations is redistributed  and the 
"center of mass" is shifted to lower energies (see Fig.~\ref{fig:fig3}). The 
$d_{xy}$ states are expected to be the least sensitive to crystal field 
changes~\cite{MorettiNJP11}. Therefore,  it is conceivable that the  $d_{xz,yz}$ 
and  $d_{z^2}$ states are shifting to lower energies. Again from DFT 
calculations (see \hyperref[appendixC]{Appendix C}), we expect the $d_{z^2}$ 
states to appear above those of $d_{xz,yz}$. Our experimental results thus 
(Fig.~\ref{fig:fig3}) suggest that the crystal field splitting $E_{{z^2}}$ in 
doped LSCO $x=0.12$ is smaller compared to LCO. Yet, the zone-boundary 
dispersion is larger in LSCO $x=0.12$ (Fig.~\ref{fig:fig2}). The present 
experiment is therefore not lending support for a correlation between the zone-boundary
dispersion and the crystal field splitting $E_{{z^2}}$.\\[1mm]

The spin-excitation dispersion of doped LSCO is analyzed using an effective 
Heisenberg Hamiltonian derived from a $t-t^{\prime}-t^{\prime\prime}-U$ Hubbard 
model~\cite{DelannoyPRB2009,HeadingsPRL10,DallaPiazzaPRB12}. This discussion has 
three steps. First, an approximative analytical expression for the zone-boundary 
dispersion is derived. Next, we compare to the experimentally obtained results 
using the known single-band tight-binding values of $t, t^{\prime}$ and 
$t^{\prime\prime}$. It is shown that this approach leads to unrealistically 
low values of the Coulomb interaction $U$. The $d_{z^2}$ band is therefore 
included. This two-orbital scenario allows us to describe the zone boundary 
dispersion with more realistic input parameters, as presented in the last part of the 
discussion. \\[1mm]

The simplest version of the Hubbard model contains only three parameters:  the 
Coulomb interaction $U$, the band width ($4t$), and a renormalization factor $Z$ 
-- known to have little momentum dependence. To lowest order in $J_1=4t^2/U$, no 
magnon dispersion is expected along the zone boundary. Therefore, to explain the 
zone-boundary dispersion -- first observed on La$_2$CuO$_4$ -- higher order 
terms $J_2=4t^4/U^3$ were included~\cite{ColdeaPRL01,HeadingsPRL10} to the 
model. Later, it has been pointed out that higher-order hopping terms 
$t^{\prime}$ and $t^{\prime\prime}$ can also contribute 
significantly~\cite{DelannoyPRB2009,DallaPiazzaPRB12}. Generally, the effective 
Heisenberg model yields a dispersion~\cite{DelannoyPRB2009,DallaPiazzaPRB12} 
$\omega(\overline{q})= Z \sqrt{A(\overline{q})^2-B(\overline{q})^2}$ where 
$A(\overline{q})$ and $B(\overline{q})$ -- given in the \hyperref[appendixA]{Appendix A} -- are 
depending on $U, t, t^{\prime}$ and $t^{\prime\prime}$. The zone boundary 
dispersion can be quantified by $E_{ZB}=\omega(\nicefrac{1}{2},0) - 
\omega(\nicefrac{1}{4},\nicefrac{1}{4})$. Using the single-band Hubbard model 
with realistic parameters~\cite{YoshidaPRB06,PavariniPRL01,DelannoyPRB2009} 
($U/t\sim8$, $\lvert t^{\prime} \rvert \leq t/2$ and 
$t^{\prime\prime}=-t^{\prime}/2$) for hole doped cuprates, we find (see 
\hyperref[appendixA]{Appendix A}):
\begin{equation} 
\frac{E_{ZB}}{12ZJ_2}\approx 
1+\frac{1}{12}\left[112-\left(\frac{U}{t}\right)^2\right]\left(\frac{t^ { \prime 
} } { t }\right)^2.
\end{equation}
A key prediction is thus that  $E_{ZB}$ scales as $(t^{\prime}/t)^2$ with a 
pre-factor that depends on $(U/t)^2$. \\[1mm]
    
\begin{figure}
	\begin{center}
		\includegraphics[width=0.4\textwidth]{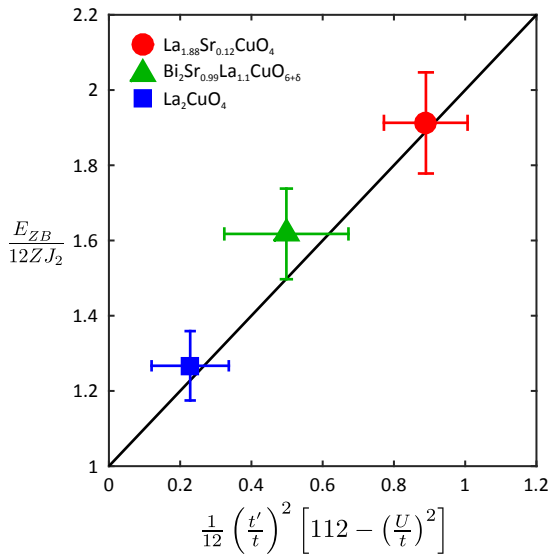}
	\end{center}
	\caption{(Color online) Experimentally obtained zone-boundary dispersion 
$E_{ZB}$, normalized to $12 Z J_2$ -- the expected theoretical value for 
$t^{\prime}=t^{\prime\prime}=0$. To obtain $J_2$ the spin-excitation dispersion 
is fitted with $U/t$ and $t^{\prime}/t$ as open parameters while keeping 
$t=0.43$~eV, $t^{\prime\prime}=-t^{\prime}/2$ and $Z=1.219$ fixed. Data points 
are obtained from fitting data on LSCO presented here (red circle) along with 
already published spin-excitation dispersions on 
LCO~\cite{ColdeaPRL01,HeadingsPRL10} (blue square) and Bi$2201$~\cite{Peng16} 
(green triangle). Error bars stem from the standard deviations of the fitting 
parameters $U/t$ and $t^{\prime}/t$. The solid line is the predicted dependence 
of the $t-t^{\prime}-t^{\prime\prime}-U$ Hubbard model with $U/t=8$.}
	\label{fig:fig4}
\end{figure}

This effective Heisenberg model is in principle not applicable to doped and 
hence antiferromagnetically disordered cuprates. For an exact description of the 
data, more sophisticated numerical methods have been 
developed~\cite{JiaNATC2014}. However, in the absence of  analytical models, the 
Heisenberg model serves as a useful effective parametrization tool to describe 
the damped spin excitations. Within a single-band tight-binding model, angle 
resolved photoemission spectroscopy (ARPES) experiments have found that 
$t^{\prime}$ decreases slightly with increasing 	
doping~\cite{YoshidaPRB06,ChangPRB08a}. The stronger zone-boundary dispersion 
can thus not be attributed to an increase of $t^{\prime}$. Parameterizing the 
doping dependent zone-boundary dispersion would thus imply a strong 
renormalization of $U$ with increasing doping.  For example, if we set  
$4t=1720$~meV (obtained from local density approximation (LDA) and 
ARPES~\cite{PavariniPRL01,FatuzzoPRB2014,ChangNATC13}) and $t^{\prime}/t=-0.16$ 
and $t^{\prime\prime}=-t^{\prime}/2$, a fit yields  $U/t\sim5$ and $Z\sim0.7$. Although these parameters provide a satisfactory description of the 
dispersion, the values of $U$ and $Z$ are not physically meaningful. \\[1mm]

This failure combined with the observation of a reduced level splitting between 
the $d_{z^2}$ and $d_{x^2-y^2}$  states (Fig.~\ref{fig:fig3}) motivates a 
two-band model. It has been demonstrated that $d_{z^2}$ states contribute to 
effectively increase the $t^{\prime}$ hopping parameter~\cite{SakakibaraPRL10}. 
Keeping $Z=1.219$ as in 
La$_2$CuO$_4$~\cite{DelannoyPRB2009} and $t^{\prime\prime}=-t^{\prime}/2$, a 
satisfactory description (solid line in Fig.~\ref{fig:fig2}) of the 
spin-excitation dispersion is obtained for $t^{\prime}/t=-0.405$ and $U/t=6.8$.
Notice that a similar ratio of $t^{\prime}/t$ has previously been inferred from 
the rounded Fermi-surface topology of 
Tl$_2$Ba$_2$CuO$_{6+x}$~\cite{PlatePRL05,PeetsNJP07,PalczewskiPRB08}  a material for which the 
$d_{z^2}$ states are expected to be much less important~\cite{SakakibaraPRB12}. 
It could thus suggest that $t^{\prime}/t\approx-0.4$ is common to single layer 
cuprates but masked in LSCO due to the repulsion between the $d_{x^2-y^2}$  and 
$d_{z^2}$ bands that pushes the van Hove singularity close to the Fermi level 
and effectively reshapes the Fermi-surface topology~\cite{SakakibaraPRL10}. The 
more realistic values of $U$ and $Z$, suggest that -- for LSCO -- the 
two-orbital character of this system is an important ingredient to accurately 
describe the spin-excitation spectrum.    

Once having extracted $U/t$ and $t^{\prime}/t$ by fitting the experimental 
spin-excitation spectrum, we plot -- in Fig.~\ref{fig:fig4} -- the normalized 
zone-boundary dispersion $E_{ZB}/(12 Z J_2)$ versus  $\frac{1}{12} 
\left(t^{\prime}/t\right)^2 \left[112- \left( U/t \right)^2\right]$. The same 
parameters were extracted (see Table~\ref{tab:tab1} in the Appendix) from 
published RIXS data on La$_2$CuO$_4$ and 
Bi$_2$Sr$_{0.99}$La${_{1.1}}$CuO$_{6+\delta}$~\cite{Peng16} and plotted in 
Fig.~\ref{fig:fig4}. All three compounds follow approximately  the predicted 
correlation between $E_{ZB}/(12 Z J_2)$ and $\frac{1}{12} 
\left(t^{\prime}/t\right)^2  \left[112- \left( U/t \right)^2\right]$. This 
suggests that the zone-boundary dispersion is controlled by the  parameters $t^{\prime}/t$ and $U/t$. It would be interesting to extend 
this parametrization  to include higher doping concentrations of LSCO. However, 
from existing RIXS data on overdoped single crystals of LSCO it is not possible 
to perform the analysis presented here~\cite{WakimotoPRB2015,MonneyPRB16}. For 
LSCO $x=0.23$, for example, the zone-boundary dispersion has not been 
measured~\cite{MonneyPRB16}.

Finally, we notice that recent RIXS experiments on LSCO thin films using 
SrLaAlO$_4$ (SLAO) substrates found a much less pronounced softening of the 
spin-excitation dispersion around the 
$(\nicefrac{1}{4},\nicefrac{1}{4})$ point~\cite{Meyers16}. A possible 
explanation is that LSCO films on SLAO have a larger $c$-axis lattice parameter 
and hence also a larger copper to apical-oxygen distance than what is found in 
bulk crystals~\cite{LocquetNat98,AbrechtPRL03}. As a consequence, the  $d_{z^2}$ 
states are less relevant and hence lead to a less pronounced zone-boundary dispersion.

\section{Conclusions}
In summary, a comprehensive RIXS study of underdoped LSCO $x=0.12$ and $0.145$ 
were presented. The spin-excitation dispersion was studied along three 
high-symmetry directions and a strong zone-boundary dispersion is reported.  The 
spin-excitation dispersion was parametrized and discussed using a Heisenberg 
Hamiltonian derived from a Hubbard model including higher-order hopping 
integrals. Within this model, the zone-boundary dispersion scales with next-nearest-neighbor 
hopping integral $t^{\prime2}$. We argue that hybridization 
between $d_{z^2}$ and $d_{x^2-y^2}$, which is especially strong in LSCO, leads 
to an enhanced $t^{\prime}$. This effect -- consistent with the observations -- leads to a 
stronger zone-boundary dispersion within the $t-t^{\prime}-t^{\prime\prime}-U$ 
Hubbard model. 

\section{Acknowledgments:}
We acknowledge support by the Swiss National Science Foundation under grant 
No. BSSGI0$\_155873$ and through the SINERGIA network Mott Physics Beyond 
the Heisenberg Model. This work was performed at the ADRESS beamline of the SLS 
at the Paul Scherrer Institut, Villigen PSI, Switzerland. We thank the ADRESS 
beamline staff for technical support. M.D. and T.S. have been partially funded 
by the Swiss National Science Foundation within the D-A-CH program (SNSF 
Research Grant No. $200021$L $141325$). X.L. acknowledges financial support from the 
European Community’s Seventh Framework Program (FP$7/2007-2013$) under grant 
agreement No. $290605$ (COFUND: PSI-FELLOW). D.M.N. was supported by the Swiss 
National Science Foundation under
Grant No. $51$NF$40$\_$141828$ through the NCCR-MARVEL. C.M. acknowledges support
from the Swiss National Science Foundation under grant $PZ00P2\_ 154867$.

\newpage
\section{Appendix A}\label{appendixA}
Here we describe the spin-excitation dispersion of the 
Heisenberg Hamiltonian derived from the $t-t^{\prime}-t^{\prime\prime}-U$ 
Hubbard model in two steps. We first consider the simplest model where 
$t^{\prime}=t^{\prime\prime}=0$ before including higher-order hopping terms. 

Generally the dispersion takes the form:
\begin{equation*}
\omega(\overline{q})= Z \sqrt{A(\overline{q})^2-B(\overline{q})^2}
\end{equation*}
where $Z$ is a renormalization factor and $\overline{q}=(h,k)$. When the Hubbard model contains only the 
nearest-neighbor hopping  integral $t$, we expand $A(\overline{q})$ and $B(\overline{q})$ to second
order in $t$:
\begin{equation}
A(\overline{q})= A_0+A_1+\dots \quad \&  \quad B(\overline{q})= B_0+B_1+\dots
\end{equation}
To express $A_i$ and $B_i$, we define: $J_1= \frac{4t^2}{U}$ and $J_2= 
\frac{4t^4}{U^3}$. Moreover we set:
\begin{equation}
P_j(h,k)= \cos { j ha}+\cos { j ka}
\end{equation}
\begin{equation}
X_j(h,k)= \cos { j h a}\cdot\cos { j k a}
\end{equation}
\begin{equation}
X_{3a}(h,k)= \cos {3  ha}\cdot \cos { ka} +\cos {ha}\cdot \cos 
{3  ka},
\end{equation}
where $j=1,2,3,$ or $4$. With this notation we have:
\begin{equation}
A_0= 2 J_1 \quad \&  \quad B_0= - J_1  P_1 
\end{equation}
and 
\begin{equation}
A_1= J_2 \left(-26 - 8 X_1 +P_2\right)  \quad \&  \quad B_1=  16  J_2   P_1
\end{equation}

When the zone-boundary dispersion is defined by 
$E_{ZB}=\omega(\nicefrac{1}{2},0)-\omega(\nicefrac{1}{4},\nicefrac{1}{4})$, one 
finds $E_{ZB}=12ZJ_2$. Therefore, a zone-boundary dispersion is only found when 
second-order terms $J_2$ are included. Notice also that since 
$P_1(\nicefrac{1}{2},0)=P_1(\nicefrac{1}{4},\nicefrac{1}{4})=0$, the $B$-terms 
are not contributing to the zone-boundary dispersion.

\begin{figure}[]
	\begin{center}
		\includegraphics[width=0.4\textwidth]{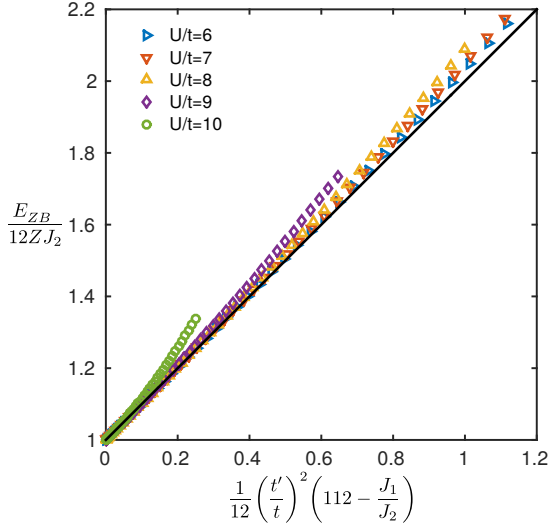}
	\end{center}
	\caption{(Color online) zone-boundary dispersion $E_{ZB}$ normalized to 
$12ZJ_2$ and plotted versus $\frac{1}{12}\left(112-J_1/J_2 
\right)(t^{\prime}/t)^2$. Data points are exact  numerical solutions of the 
Hubbard model for values several of $U/t$ (as indicated) and 
$t^{\prime\prime}=-t^{\prime}/2$. The solid line is the approximated analytical 
solution  for $U/t=8$.} \label{fig:fig1_suppl}
\end{figure}
\begin{figure}[b]
	\begin{center}
		\includegraphics[width=0.4\textwidth]{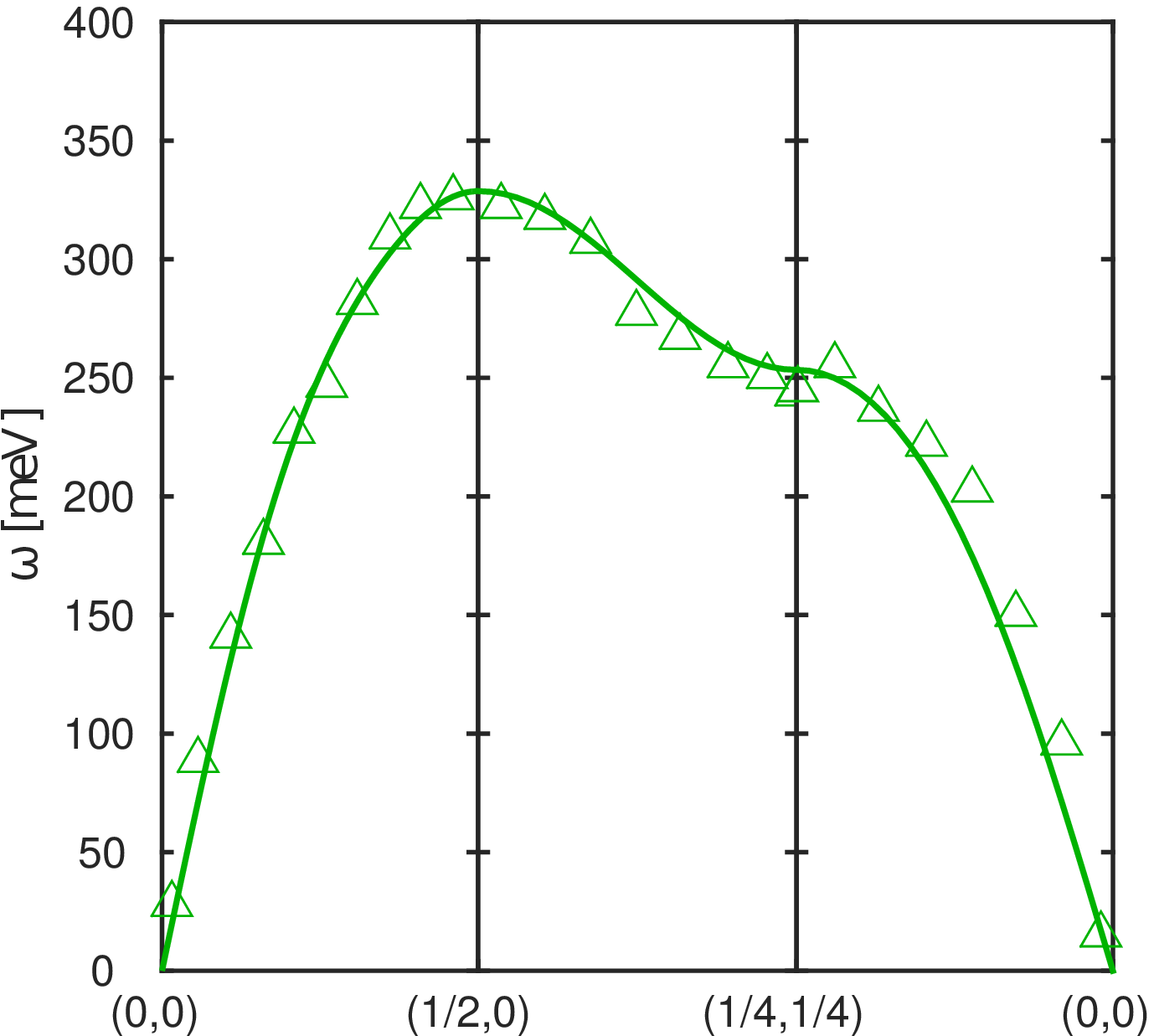}
	\end{center}
	\caption{Spin-excitation spectrum of Bi$2201$ from 
Ref.~\onlinecite{Peng16}. The solid line is a fit to the 
$t-t^{\prime}-t^{\prime\prime}-U$ Hubbard model.} \label{fig:fig2_suppl}
\end{figure}

\begin{figure*}[]
	\begin{center}
		\includegraphics[width=0.995\textwidth]{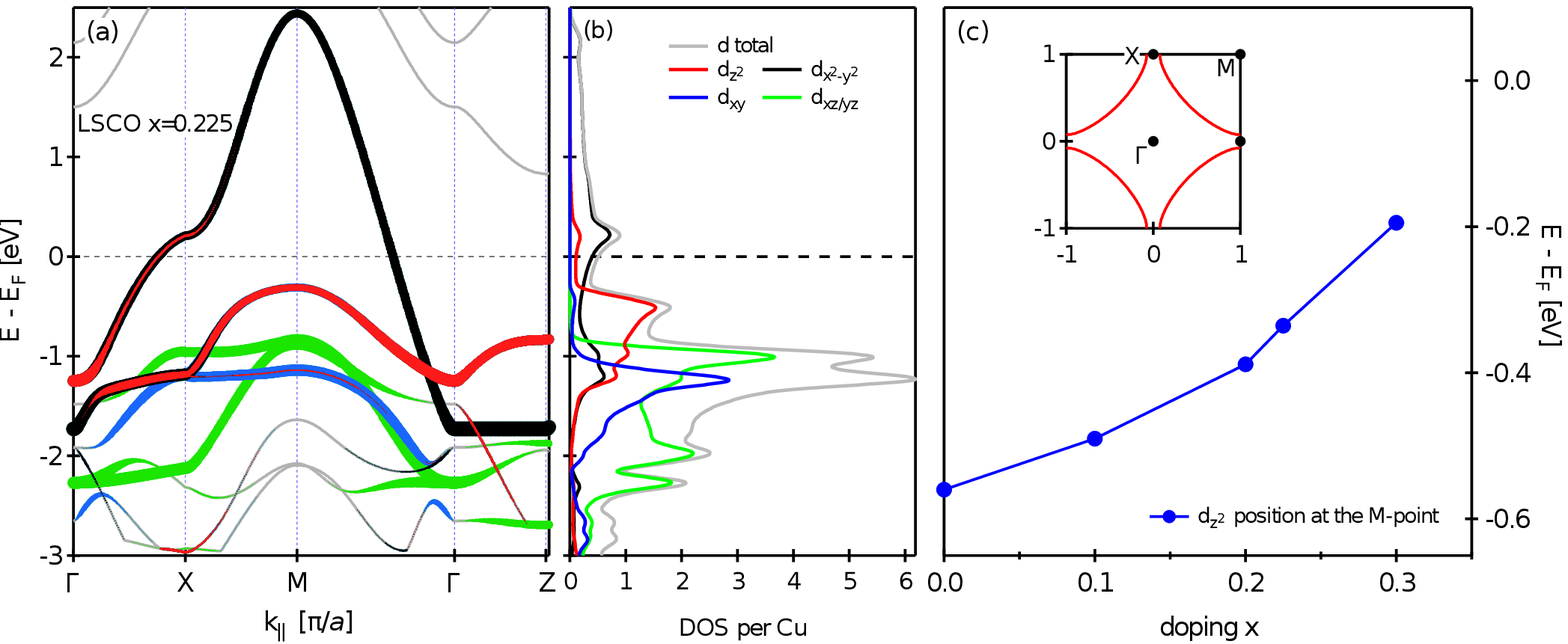}
	\end{center}
	\caption{(Color online) Density functional theory calculations of 
La$_{2-x}$Sr$_x$CuO$_4$. (a) calculated bandstructure along high symmetry 
directions (see inset of panel(c)) in the tetragonal crystal structure for 
$x=0.225$ (Ref.~\onlinecite{RadaelliPRB94}). (b) density of states derived by 
the different Cu $3d$ orbitals. The electronic structure has been shifted such  
that the overall $3d$-shell filling reflects the doping $x$. (c) doping 
dependence of the $d_{z^2}$ band derived at the M point.} \label{fig:fig3_suppl}
\end{figure*}

Now let us include second-nearest $t^{\prime}$ and third-nearest-neighbor 
$t^{\prime\prime}$ hopping integrals. This involves several additional 
contributions to $A(\overline{q})$ and $B(\overline{q})$:  
\begin{equation}
A(\overline{q})= A_0+A_1+A^{\prime}_0 +A^{\prime\prime}_0+A^{\prime}_c+A^{\prime}_1 
+A^{\prime\prime}_c+A^{\prime\prime}_1
\end{equation}
\begin{equation}
B(\overline{q})= B_0+B_1+B^{\prime}_c.
\end{equation}
To express these new terms, we introduce the following notation $J^{\prime}_1= 
\frac{4t^{\prime2}}{U}$,  $J^{\prime}_2= \frac{4t^{\prime4}}{U^3}$, 
$J^{\prime\prime}_1= \frac{4t^{\prime\prime2}}{U}$ and $J^{\prime\prime}_2= 
\frac{4t^{\prime\prime4}}{U^3}$. Geometrically the following contributions 
correspond to different hopping path combinations including the cyclic ones. 
\begin{equation}\label{eq:s1}
A^{\prime}_0= 2J^{\prime}_1 (X_1-1)  \quad \&  \quad A^{\prime\prime}_0=  
J^{\prime\prime}_1 (P_2-2)
\end{equation}
\begin{equation}\label{eq:s2}
A^{\prime}_c=  -\frac{8 J_1}{U^2} \left( -  t^{\prime\,2}+ 4 t^{\prime} 
t^{\prime\prime\,  } -2\, t^{\prime\prime\,2} \right) (P_2-2) 
\end{equation}
\begin{equation}
B^{\prime}_c= -\frac{4 J_1}{U^2}\left[ \left(6\, t^{\prime\,  2} -4\,  
t^{\prime} t^{\prime\prime} \right)  (X_1-1) 
+3\,  t^{\prime\prime\,  2}  ( P_2-2) \right] P_1
\end{equation}
\begin{equation}
A^{\prime}_1=2J^{\prime}_2   (X_2+4 X_1-2 P_2-1) 
\end{equation}
\begin{equation}
A^{\prime\prime}_c=  \frac{2 J^{\prime}_1  J^{\prime\prime}_1}{U} 
\left(-3 X_2+2 X_1 + 5 P_2 -X_{3a}  - 7 \right)
\end{equation}
\begin{equation}
A^{\prime\prime}_1=  J^{\prime\prime}_2  (P_4-8 X_2+4 P_2-2)
\end{equation}

\begin{table}[t]
\begin{center}
\caption{Parametrization -- using the Hubbard model -- of the spin-excitation 
dispersion of LCO~\cite{ColdeaPRL01,HeadingsPRL10}, LSCO $x=0.12$ (this work) 
and Bi$2201$~\cite{Peng16}. (\textborn) Values obtained from the fit using the 
same procedure as described in Appendix B, which thus can be directly 
compared.}
\label{tab:tab1}
\begin{ruledtabular}
\begin{tabular}{ccccccc}
La$_{2-x}$Sr$_x$CuO$_4$&  $U$ (eV) & $U/t$  & $t^{\prime}/t$ & 
$t^{\prime\prime}/t$ & $Z$ & Ref.\\ \hline
$x=0$ &  2.2 & 7.4 & 0 & 0 & $ 1.18 $ &~\onlinecite{ColdeaPRL01,HeadingsPRL10} 
\\
$x=0$ &  3.6 & 8.3  & -0.313 & 0.167 & 1.219 &~\onlinecite{DelannoyPRB2009} \\
$x=0$ &  3.9 & 9.1  & -0.308 & 0.154 & 1.219  & \textborn \\
$x=0.12$ &  2.9 & 6.8 & -0.405 & 0.202 & 1.219 & \textborn \\\hline
Bi$2201$ \\\hline
$x=0$ &  3.4 & 8.0  & -0.352 & 0.176 & 1.219  & \textborn \\
\end{tabular}
\end{ruledtabular}
\end{center}	
\end{table}

As $B^{\prime}_c$ scales with $P_1$, it is again found that $B(\overline{q})$ does not 
contribute to the zone-boundary dispersion. In Fig.~\ref{fig:fig1_suppl}, we 
show the numerical evaluation of $E_{ZB}$ for realistic values of $U/t$, 
$t^{\prime}/t$ and with $t^{\prime\prime}=-t^{\prime}/2$. When neglecting terms 
scaling with $J^{\prime}_2$, $J^{\prime\prime}_2$ and $J^{\prime}_1 
J^{\prime\prime}_1$, only Eq.~\ref{eq:s1} and \ref{eq:s2} contribute. Using 
$P_2(\nicefrac{1}{2},0)=2$, $P_2(\nicefrac{1}{4},\nicefrac{1}{4})=-2$, 
$X_1(\nicefrac{1}{2},0)=-1$ and $X_1(\nicefrac{1}{4},\nicefrac{1}{4})=0$, we 
find:
\begin{equation}
\frac{E_{ZB}}{12ZJ_2}\approx 
1+\frac{1}{12}\left(112-\frac{J_1}{J_2}\right)\left(\frac{t^{\prime}}{t}
\right)^2.
\end{equation}
This approximation  is valid as long as:
\begin{equation}
\frac{U}{t}\geq \sqrt{\frac{28+112 \left( \frac{t^{\prime}}{t} 
\right)^2}{2+3 \left( \frac{t^{\prime}}{t} \right)^2}}, \newline
\quad \textrm{and} \quad \left|{\frac{t^{\prime}}{t}}\right|  \lesssim  0.686.
\end{equation}

As shown in Fig.~\ref{fig:fig1_suppl}, this analytical expression is a good 
approximation to the full numerical calculation. Thus it is justified to neglect 
terms scaling with $J^{\prime}_2$, $J^{\prime\prime}_2$ and $J^{\prime}_1 
J^{\prime\prime}_1$ for a realistic cuprate values of $U/t$ and $t^{\prime}/t$.

\section{Appendix B}
Now, having derived the spin-excitation dispersion within the 
$t-t^{\prime}-t^{\prime\prime}-U$ Hubbard model, it is possible to fit the 
experimentally observed dispersion. A final comment goes to the prefactor $Z$. 
It is found that, including higher-order hopping integrals $t^{\prime}$ and 
$t^{\prime\prime}$, $Z$ has a slowly varying momentum dependence. To simplify 
our analysis we used the mean value obtained~\cite{DelannoyPRB2009} in the first 
Brillouin zone for the half filled compound La$_2$CuO$_4$. We thus have 
$Z=1.219$ constant. From ARPES~\cite{FatuzzoPRB2014,ChangNATC13} experiments and 
LDA calculations~\cite{PavariniPRL01} we have that $t=0.43$~eV and 
$t^{\prime\prime}=-t^{\prime}/2$. Our fitting parameters are thus $U$ and 
$t^{\prime}$. In this fashion we obtain a good description of the 
spin-excitation dispersion of LCO and LSCO $x=0.12$ (see Fig.~\ref{fig:fig2} in 
the main text). The obtained values are given in Table~\ref{tab:tab1}. In 
Fig.~\ref{fig:fig2_suppl} and Table~\ref{tab:tab1}, we display in addition  our 
fit and associated fit parameters from the spin-excitation spectrum measured on 
Bi$2201$ (Ref.~\onlinecite{Peng16}). With these values of $U$ and $t^{\prime}$, 
the relation  -- shown in Fig.~\ref{fig:fig4} -- between $E_{ZB}$ and 
$t^{\prime}$ is established.

\section{Appendix C}\label{appendixC}
To guide our intuition of how the $d_{z^2}$ states evolve as a function of 
doping, we have carried out DTF calculations of the 
LSCO band structure as a function of doping. These calculations were performed 
using the WIEN2K package~\cite{Blaha2001} in the LTO crystal structure. The 
doping dependence of the electronic structure for LSCO was approximated by a 
rigid band shift of all Cu $d$ orbitals in order to obtain the correct 
$d$-shell filling. For every calculated doping value, the experimentally derived 
crystal structure has been used~\cite{RadaelliPRB94}. In the calculation, the 
Kohn-Sham equation is solved self-consistently by using a full-potential linear 
augmented plane wave (LAPW) method on a uniform grid of $12\times12\times12$ 
$k$ points in the Brillouin zone. The exchange-correlation term is treated 
within the generalized gradient approximation (GGA) in the parametrization of 
Perdew, Burke and Enzerhof (PBE)~\cite{PerdewPRL1996}. The plane-wave cutoff 
condition was set to $R K_{max}=7$ where $R$ is the radius of the smallest LAPW 
sphere ($i.e.$ $1.63$ bohrs) and $K_{max}$ denotes the plane-wave cutoff. 
Fig.~\ref{fig:fig3_suppl} shows the orbital and atomic resolved band structure 
and density of states (DOS) of LSCO in the tetragonal crystal structure. As 
shown in panel (a), the $d_{z^2}$ derived band disperses in a binding energy 
range of $E-E_F=-1.3~eV$ close to $\Gamma$ and $E-E_F=-0.3~eV$ at $M$. The 
orbital resolved DOS of the $d_{z^2}$ band has a peak at $E-E_F=-0.5~eV$, while 
closer to $E_F$ the $d_{z^2}$ DOS rapidly decays. This peak originates from 
the flat shape of the $d_{z^2}$ band close to $M$. Therefore to track the doping 
dependence of the $d_{z^2}$ energy level, the position of the band at the $M$ 
point is plotted as a function of doping $x$ in Fig.~\ref{fig:fig3_suppl}(c). 
With increasing doping $x$ the $d_{z^2}$ energy level approaches the Fermi 
energy. Note that our DFT calculation agrees with recently published results 
obtained by \textit{ab initio} calculations~\cite{SakakibaraPRL10}.

\bibliography{RIXS_LSCO}

\begin{thebibliography}{48}%
\makeatletter
\providecommand \@ifxundefined [1]{%
 \@ifx{#1\undefined}
}%
\providecommand \@ifnum [1]{%
 \ifnum #1\expandafter \@firstoftwo
 \else \expandafter \@secondoftwo
 \fi
}%
\providecommand \@ifx [1]{%
 \ifx #1\expandafter \@firstoftwo
 \else \expandafter \@secondoftwo
 \fi
}%
\providecommand \natexlab [1]{#1}%
\providecommand \enquote  [1]{``#1''}%
\providecommand \bibnamefont  [1]{#1}%
\providecommand \bibfnamefont [1]{#1}%
\providecommand \citenamefont [1]{#1}%
\providecommand \href@noop [0]{\@secondoftwo}%
\providecommand \href [0]{\begingroup \@sanitize@url \@href}%
\providecommand \@href[1]{\@@startlink{#1}\@@href}%
\providecommand \@@href[1]{\endgroup#1\@@endlink}%
\providecommand \@sanitize@url [0]{\catcode `\\12\catcode `\$12\catcode
  `\&12\catcode `\#12\catcode `\^12\catcode `\_12\catcode `\%12\relax}%
\providecommand \@@startlink[1]{}%
\providecommand \@@endlink[0]{}%
\providecommand \url  [0]{\begingroup\@sanitize@url \@url }%
\providecommand \@url [1]{\endgroup\@href {#1}{\urlprefix }}%
\providecommand \urlprefix  [0]{URL }%
\providecommand \Eprint [0]{\href }%
\providecommand \doibase [0]{http://dx.doi.org/}%
\providecommand \selectlanguage [0]{\@gobble}%
\providecommand \bibinfo  [0]{\@secondoftwo}%
\providecommand \bibfield  [0]{\@secondoftwo}%
\providecommand \translation [1]{[#1]}%
\providecommand \BibitemOpen [0]{}%
\providecommand \bibitemStop [0]{}%
\providecommand \bibitemNoStop [0]{.\EOS\space}%
\providecommand \EOS [0]{\spacefactor3000\relax}%
\providecommand \BibitemShut  [1]{\csname bibitem#1\endcsname}%
\let\auto@bib@innerbib\@empty
\bibitem [{\citenamefont {Lee}\ \emph {et~al.}(2006)\citenamefont {Lee},
  \citenamefont {Nagaosa},\ and\ \citenamefont {Wen}}]{LeeRMP06}%
  \BibitemOpen
  \bibfield  {author} {\bibinfo {author} {\bibfnamefont {P.~A.}\ \bibnamefont
  {Lee}}, \bibinfo {author} {\bibfnamefont {N.}~\bibnamefont {Nagaosa}}, \ and\
  \bibinfo {author} {\bibfnamefont {X.-G.}\ \bibnamefont {Wen}},\ }\href
  {\doibase 10.1103/RevModPhys.78.17} {\bibfield  {journal} {\bibinfo
  {journal} {Rev. Mod. Phys.}\ }\textbf {\bibinfo {volume} {78}},\ \bibinfo
  {pages} {17} (\bibinfo {year} {2006})}\BibitemShut {NoStop}%
\bibitem [{\citenamefont {Norman}\ \emph {et~al.}(2005)\citenamefont {Norman},
  \citenamefont {Pines},\ and\ \citenamefont {Kallin}}]{NormanAP05}%
  \BibitemOpen
  \bibfield  {author} {\bibinfo {author} {\bibfnamefont {M.~R.}\ \bibnamefont
  {Norman}}, \bibinfo {author} {\bibfnamefont {D.}~\bibnamefont {Pines}}, \
  and\ \bibinfo {author} {\bibfnamefont {C.}~\bibnamefont {Kallin}},\ }\href
  {\doibase 10.1080/00018730500459906} {\bibfield  {journal} {\bibinfo
  {journal} {Advances in Physics}\ }\textbf {\bibinfo {volume} {54}},\ \bibinfo
  {pages} {715} (\bibinfo {year} {2005})}\BibitemShut {NoStop}%
\bibitem [{\citenamefont {Ament}\ \emph {et~al.}(2011)\citenamefont {Ament},
  \citenamefont {van Veenendaal}, \citenamefont {Devereaux}, \citenamefont
  {Hill},\ and\ \citenamefont {van~den Brink}}]{AmentRMP11}%
  \BibitemOpen
  \bibfield  {author} {\bibinfo {author} {\bibfnamefont {L.~J.~P.}\
  \bibnamefont {Ament}}, \bibinfo {author} {\bibfnamefont {M.}~\bibnamefont
  {van Veenendaal}}, \bibinfo {author} {\bibfnamefont {T.~P.}\ \bibnamefont
  {Devereaux}}, \bibinfo {author} {\bibfnamefont {J.~P.}\ \bibnamefont {Hill}},
  \ and\ \bibinfo {author} {\bibfnamefont {J.}~\bibnamefont {van~den Brink}},\
  }\href {\doibase 10.1103/RevModPhys.83.705} {\bibfield  {journal} {\bibinfo
  {journal} {Rev. Mod. Phys.}\ }\textbf {\bibinfo {volume} {83}},\ \bibinfo
  {pages} {705} (\bibinfo {year} {2011})}\BibitemShut {NoStop}%
\bibitem [{\citenamefont {Tacon}\ \emph {et~al.}(2011)\citenamefont {Tacon},
  \citenamefont {Ghiringhelli}, \citenamefont {Chaloupka}, \citenamefont
  {Sala}, \citenamefont {Hinkov}, \citenamefont {Haverkort}, \citenamefont
  {Minola}, \citenamefont {Bakr}, \citenamefont {Zhou}, \citenamefont
  {Blanco-Canosa}, \citenamefont {Monney}, \citenamefont {Song}, \citenamefont
  {Sun}, \citenamefont {Lin}, \citenamefont {Luca}, \citenamefont {Salluzzo},
  \citenamefont {Khaliullin}, \citenamefont {Schmitt}, \citenamefont
  {Braicovich},\ and\ \citenamefont {Keimer}}]{TaconNATP11}%
  \BibitemOpen
  \bibfield  {author} {\bibinfo {author} {\bibfnamefont {M.~L.}\ \bibnamefont
  {Tacon}}, \bibinfo {author} {\bibfnamefont {G.}~\bibnamefont {Ghiringhelli}},
  \bibinfo {author} {\bibfnamefont {J.}~\bibnamefont {Chaloupka}}, \bibinfo
  {author} {\bibfnamefont {M.~M.}\ \bibnamefont {Sala}}, \bibinfo {author}
  {\bibfnamefont {V.}~\bibnamefont {Hinkov}}, \bibinfo {author} {\bibfnamefont
  {M.~W.}\ \bibnamefont {Haverkort}}, \bibinfo {author} {\bibfnamefont
  {M.}~\bibnamefont {Minola}}, \bibinfo {author} {\bibfnamefont
  {M.}~\bibnamefont {Bakr}}, \bibinfo {author} {\bibfnamefont {K.~J.}\
  \bibnamefont {Zhou}}, \bibinfo {author} {\bibfnamefont {S.}~\bibnamefont
  {Blanco-Canosa}}, \bibinfo {author} {\bibfnamefont {C.}~\bibnamefont
  {Monney}}, \bibinfo {author} {\bibfnamefont {Y.~T.}\ \bibnamefont {Song}},
  \bibinfo {author} {\bibfnamefont {G.~L.}\ \bibnamefont {Sun}}, \bibinfo
  {author} {\bibfnamefont {C.~T.}\ \bibnamefont {Lin}}, \bibinfo {author}
  {\bibfnamefont {G.~M.~D.}\ \bibnamefont {Luca}}, \bibinfo {author}
  {\bibfnamefont {M.}~\bibnamefont {Salluzzo}}, \bibinfo {author}
  {\bibfnamefont {G.}~\bibnamefont {Khaliullin}}, \bibinfo {author}
  {\bibfnamefont {T.}~\bibnamefont {Schmitt}}, \bibinfo {author} {\bibfnamefont
  {L.}~\bibnamefont {Braicovich}}, \ and\ \bibinfo {author} {\bibfnamefont
  {B.}~\bibnamefont {Keimer}},\ }\href {\doibase 10.1038/nphys2041} {\bibfield
  {journal} {\bibinfo  {journal} {Nat. Phys.}\ }\textbf {\bibinfo {volume}
  {7}},\ \bibinfo {pages} {725} (\bibinfo {year} {2011})}\BibitemShut {NoStop}%
\bibitem [{\citenamefont {{M. P. M. Dean}}\ \emph {et~al.}(2013)\citenamefont
  {{M. P. M. Dean}}, \citenamefont {{G. Dellea}}, \citenamefont {{R. S.
  Springell}}, \citenamefont {{F. Yakhou-Harris}}, \citenamefont {{K. Kummer}},
  \citenamefont {{N. B. Brookes}}, \citenamefont {{X. Liu}}, \citenamefont
  {{Y-J. Sun}}, \citenamefont {{J. Strle}}, \citenamefont {{T. Schmitt}},
  \citenamefont {{L. Braicovich}}, \citenamefont {{G. Ghiringhelli}},
  \citenamefont {{I. Bo\v{z}ovi{\'c}}},\ and\ \citenamefont {{J. P.
  Hill}}}]{DeanNATM13}%
  \BibitemOpen
  \bibfield  {author} {\bibinfo {author} {\bibnamefont {{M. P. M. Dean}}},
  \bibinfo {author} {\bibnamefont {{G. Dellea}}}, \bibinfo {author}
  {\bibnamefont {{R. S. Springell}}}, \bibinfo {author} {\bibnamefont {{F.
  Yakhou-Harris}}}, \bibinfo {author} {\bibnamefont {{K. Kummer}}}, \bibinfo
  {author} {\bibnamefont {{N. B. Brookes}}}, \bibinfo {author} {\bibnamefont
  {{X. Liu}}}, \bibinfo {author} {\bibnamefont {{Y-J. Sun}}}, \bibinfo {author}
  {\bibnamefont {{J. Strle}}}, \bibinfo {author} {\bibnamefont {{T. Schmitt}}},
  \bibinfo {author} {\bibnamefont {{L. Braicovich}}}, \bibinfo {author}
  {\bibnamefont {{G. Ghiringhelli}}}, \bibinfo {author} {\bibnamefont {{I.
  Bo\v{z}ovi{\'c}}}}, \ and\ \bibinfo {author} {\bibnamefont {{J. P. Hill}}},\
  }\href {\doibase 10.1038/nmat3723} {\bibfield  {journal} {\bibinfo  {journal}
  {Nat Mater}\ }\textbf {\bibinfo {volume} {12}},\ \bibinfo {pages} {1019}
  (\bibinfo {year} {2013})}\BibitemShut {NoStop}%
\bibitem [{\citenamefont {Coldea}\ \emph {et~al.}(2001)\citenamefont {Coldea},
  \citenamefont {Hayden}, \citenamefont {Aeppli}, \citenamefont {Perring},
  \citenamefont {Frost}, \citenamefont {Mason}, \citenamefont {Cheong},\ and\
  \citenamefont {Fisk}}]{ColdeaPRL01}%
  \BibitemOpen
  \bibfield  {author} {\bibinfo {author} {\bibfnamefont {R.}~\bibnamefont
  {Coldea}}, \bibinfo {author} {\bibfnamefont {S.~M.}\ \bibnamefont {Hayden}},
  \bibinfo {author} {\bibfnamefont {G.}~\bibnamefont {Aeppli}}, \bibinfo
  {author} {\bibfnamefont {T.~G.}\ \bibnamefont {Perring}}, \bibinfo {author}
  {\bibfnamefont {C.~D.}\ \bibnamefont {Frost}}, \bibinfo {author}
  {\bibfnamefont {T.~E.}\ \bibnamefont {Mason}}, \bibinfo {author}
  {\bibfnamefont {S.-W.}\ \bibnamefont {Cheong}}, \ and\ \bibinfo {author}
  {\bibfnamefont {Z.}~\bibnamefont {Fisk}},\ }\href {\doibase
  10.1103/PhysRevLett.86.5377} {\bibfield  {journal} {\bibinfo  {journal}
  {Phys. Rev. Lett.}\ }\textbf {\bibinfo {volume} {86}},\ \bibinfo {pages}
  {5377} (\bibinfo {year} {2001})}\BibitemShut {NoStop}%
\bibitem [{\citenamefont {Headings}\ \emph {et~al.}(2010)\citenamefont
  {Headings}, \citenamefont {Hayden}, \citenamefont {Coldea},\ and\
  \citenamefont {Perring}}]{HeadingsPRL10}%
  \BibitemOpen
  \bibfield  {author} {\bibinfo {author} {\bibfnamefont {N.~S.}\ \bibnamefont
  {Headings}}, \bibinfo {author} {\bibfnamefont {S.~M.}\ \bibnamefont
  {Hayden}}, \bibinfo {author} {\bibfnamefont {R.}~\bibnamefont {Coldea}}, \
  and\ \bibinfo {author} {\bibfnamefont {T.~G.}\ \bibnamefont {Perring}},\
  }\href {\doibase 10.1103/PhysRevLett.105.247001} {\bibfield  {journal}
  {\bibinfo  {journal} {Phys. Rev. Lett.}\ }\textbf {\bibinfo {volume} {105}},\
  \bibinfo {pages} {247001} (\bibinfo {year} {2010})}\BibitemShut {NoStop}%
\bibitem [{\citenamefont {Delannoy}\ \emph {et~al.}(2009)\citenamefont
  {Delannoy}, \citenamefont {Gingras}, \citenamefont {Holdsworth},\ and\
  \citenamefont {Tremblay}}]{DelannoyPRB2009}%
  \BibitemOpen
  \bibfield  {author} {\bibinfo {author} {\bibfnamefont {J.-Y.~P.}\
  \bibnamefont {Delannoy}}, \bibinfo {author} {\bibfnamefont {M.~J.~P.}\
  \bibnamefont {Gingras}}, \bibinfo {author} {\bibfnamefont {P.~C.~W.}\
  \bibnamefont {Holdsworth}}, \ and\ \bibinfo {author} {\bibfnamefont
  {A.-M.~S.}\ \bibnamefont {Tremblay}},\ }\href {\doibase
  10.1103/PhysRevB.79.235130} {\bibfield  {journal} {\bibinfo  {journal} {Phys.
  Rev. B}\ }\textbf {\bibinfo {volume} {79}},\ \bibinfo {pages} {235130}
  (\bibinfo {year} {2009})}\BibitemShut {NoStop}%
\bibitem [{\citenamefont {{Dalla Piazza}}\ \emph {et~al.}(2012)\citenamefont
  {{Dalla Piazza}}, \citenamefont {Mourigal}, \citenamefont {Guarise},
  \citenamefont {Berger}, \citenamefont {Schmitt}, \citenamefont {Zhou},
  \citenamefont {Grioni},\ and\ \citenamefont {R{\o}nnow}}]{DallaPiazzaPRB12}%
  \BibitemOpen
  \bibfield  {author} {\bibinfo {author} {\bibfnamefont {B.}~\bibnamefont
  {{Dalla Piazza}}}, \bibinfo {author} {\bibfnamefont {M.}~\bibnamefont
  {Mourigal}}, \bibinfo {author} {\bibfnamefont {M.}~\bibnamefont {Guarise}},
  \bibinfo {author} {\bibfnamefont {H.}~\bibnamefont {Berger}}, \bibinfo
  {author} {\bibfnamefont {T.}~\bibnamefont {Schmitt}}, \bibinfo {author}
  {\bibfnamefont {K.~J.}\ \bibnamefont {Zhou}}, \bibinfo {author}
  {\bibfnamefont {M.}~\bibnamefont {Grioni}}, \ and\ \bibinfo {author}
  {\bibfnamefont {H.~M.}\ \bibnamefont {R{\o}nnow}},\ }\href {\doibase
  10.1103/PhysRevB.85.100508} {\bibfield  {journal} {\bibinfo  {journal} {Phys.
  Rev. B}\ }\textbf {\bibinfo {volume} {85}},\ \bibinfo {pages} {100508}
  (\bibinfo {year} {2012})}\BibitemShut {NoStop}%
\bibitem [{\citenamefont {Yoshida}\ \emph {et~al.}(2006)\citenamefont
  {Yoshida}, \citenamefont {Zhou}, \citenamefont {Tanaka}, \citenamefont
  {Yang}, \citenamefont {Hussain}, \citenamefont {Shen}, \citenamefont
  {Fujimori}, \citenamefont {Sahrakorpi}, \citenamefont {Lindroos},
  \citenamefont {Markiewicz}, \citenamefont {Bansil}, \citenamefont {Komiya},
  \citenamefont {Ando}, \citenamefont {Eisaki}, \citenamefont {Kakeshita},\
  and\ \citenamefont {Uchida}}]{YoshidaPRB06}%
  \BibitemOpen
  \bibfield  {author} {\bibinfo {author} {\bibfnamefont {T.}~\bibnamefont
  {Yoshida}}, \bibinfo {author} {\bibfnamefont {X.~J.}\ \bibnamefont {Zhou}},
  \bibinfo {author} {\bibfnamefont {K.}~\bibnamefont {Tanaka}}, \bibinfo
  {author} {\bibfnamefont {W.~L.}\ \bibnamefont {Yang}}, \bibinfo {author}
  {\bibfnamefont {Z.}~\bibnamefont {Hussain}}, \bibinfo {author} {\bibfnamefont
  {Z.-X.}\ \bibnamefont {Shen}}, \bibinfo {author} {\bibfnamefont
  {A.}~\bibnamefont {Fujimori}}, \bibinfo {author} {\bibfnamefont
  {S.}~\bibnamefont {Sahrakorpi}}, \bibinfo {author} {\bibfnamefont
  {M.}~\bibnamefont {Lindroos}}, \bibinfo {author} {\bibfnamefont {R.~S.}\
  \bibnamefont {Markiewicz}}, \bibinfo {author} {\bibfnamefont
  {A.}~\bibnamefont {Bansil}}, \bibinfo {author} {\bibfnamefont
  {S.}~\bibnamefont {Komiya}}, \bibinfo {author} {\bibfnamefont
  {Y.}~\bibnamefont {Ando}}, \bibinfo {author} {\bibfnamefont {H.}~\bibnamefont
  {Eisaki}}, \bibinfo {author} {\bibfnamefont {T.}~\bibnamefont {Kakeshita}}, \
  and\ \bibinfo {author} {\bibfnamefont {S.}~\bibnamefont {Uchida}},\ }\href
  {\doibase 10.1103/PhysRevB.74.224510} {\bibfield  {journal} {\bibinfo
  {journal} {Phys. Rev. B}\ }\textbf {\bibinfo {volume} {74}},\ \bibinfo
  {pages} {224510} (\bibinfo {year} {2006})}\BibitemShut {NoStop}%
\bibitem [{\citenamefont {Pavarini}\ \emph {et~al.}(2001)\citenamefont
  {Pavarini}, \citenamefont {Dasgupta}, \citenamefont {Saha-Dasgupta},
  \citenamefont {Jepsen},\ and\ \citenamefont {Andersen}}]{PavariniPRL01}%
  \BibitemOpen
  \bibfield  {author} {\bibinfo {author} {\bibfnamefont {E.}~\bibnamefont
  {Pavarini}}, \bibinfo {author} {\bibfnamefont {I.}~\bibnamefont {Dasgupta}},
  \bibinfo {author} {\bibfnamefont {T.}~\bibnamefont {Saha-Dasgupta}}, \bibinfo
  {author} {\bibfnamefont {O.}~\bibnamefont {Jepsen}}, \ and\ \bibinfo {author}
  {\bibfnamefont {O.~K.}\ \bibnamefont {Andersen}},\ }\href {\doibase
  10.1103/PhysRevLett.87.047003} {\bibfield  {journal} {\bibinfo  {journal}
  {Phys. Rev. Lett.}\ }\textbf {\bibinfo {volume} {87}},\ \bibinfo {pages}
  {047003} (\bibinfo {year} {2001})}\BibitemShut {NoStop}%
\bibitem [{\citenamefont {White}\ and\ \citenamefont
  {Scalapino}(1999)}]{WhitePRB99}%
  \BibitemOpen
  \bibfield  {author} {\bibinfo {author} {\bibfnamefont {S.~R.}\ \bibnamefont
  {White}}\ and\ \bibinfo {author} {\bibfnamefont {D.~J.}\ \bibnamefont
  {Scalapino}},\ }\href {\doibase 10.1103/PhysRevB.60.R753} {\bibfield
  {journal} {\bibinfo  {journal} {Phys. Rev. B}\ }\textbf {\bibinfo {volume}
  {60}},\ \bibinfo {pages} {R753} (\bibinfo {year} {1999})}\BibitemShut
  {NoStop}%
\bibitem [{\citenamefont {Maier}\ \emph {et~al.}(2000)\citenamefont {Maier},
  \citenamefont {Jarrell}, \citenamefont {Pruschke},\ and\ \citenamefont
  {Keller}}]{MaierPRL00}%
  \BibitemOpen
  \bibfield  {author} {\bibinfo {author} {\bibfnamefont {T.}~\bibnamefont
  {Maier}}, \bibinfo {author} {\bibfnamefont {M.}~\bibnamefont {Jarrell}},
  \bibinfo {author} {\bibfnamefont {T.}~\bibnamefont {Pruschke}}, \ and\
  \bibinfo {author} {\bibfnamefont {J.}~\bibnamefont {Keller}},\ }\href
  {\doibase 10.1103/PhysRevLett.85.1524} {\bibfield  {journal} {\bibinfo
  {journal} {Phys. Rev. Lett.}\ }\textbf {\bibinfo {volume} {85}},\ \bibinfo
  {pages} {1524} (\bibinfo {year} {2000})}\BibitemShut {NoStop}%
\bibitem [{\citenamefont {Sakakibara}\ \emph {et~al.}(2010)\citenamefont
  {Sakakibara}, \citenamefont {Usui}, \citenamefont {Kuroki}, \citenamefont
  {Arita},\ and\ \citenamefont {Aoki}}]{SakakibaraPRL10}%
  \BibitemOpen
  \bibfield  {author} {\bibinfo {author} {\bibfnamefont {H.}~\bibnamefont
  {Sakakibara}}, \bibinfo {author} {\bibfnamefont {H.}~\bibnamefont {Usui}},
  \bibinfo {author} {\bibfnamefont {K.}~\bibnamefont {Kuroki}}, \bibinfo
  {author} {\bibfnamefont {R.}~\bibnamefont {Arita}}, \ and\ \bibinfo {author}
  {\bibfnamefont {H.}~\bibnamefont {Aoki}},\ }\href {\doibase
  10.1103/PhysRevLett.105.057003} {\bibfield  {journal} {\bibinfo  {journal}
  {Phys. Rev. Lett.}\ }\textbf {\bibinfo {volume} {105}},\ \bibinfo {pages}
  {057003} (\bibinfo {year} {2010})}\BibitemShut {NoStop}%
\bibitem [{\citenamefont {Chang}\ \emph
  {et~al.}(2008{\natexlab{a}})\citenamefont {Chang}, \citenamefont {Shi},
  \citenamefont {Pailh{\'e}s}, \citenamefont {M{\aa}nsson}, \citenamefont
  {Claesson}, \citenamefont {Tjernberg}, \citenamefont {Bendounan},
  \citenamefont {Sassa}, \citenamefont {Patthey}, \citenamefont {Momono},
  \citenamefont {Oda}, \citenamefont {Ido}, \citenamefont {Guerrero},
  \citenamefont {Mudry},\ and\ \citenamefont {Mesot}}]{ChangPRB08a}%
  \BibitemOpen
  \bibfield  {author} {\bibinfo {author} {\bibfnamefont {J.}~\bibnamefont
  {Chang}}, \bibinfo {author} {\bibfnamefont {M.}~\bibnamefont {Shi}}, \bibinfo
  {author} {\bibfnamefont {S.}~\bibnamefont {Pailh{\'e}s}}, \bibinfo {author}
  {\bibfnamefont {M.}~\bibnamefont {M{\aa}nsson}}, \bibinfo {author}
  {\bibfnamefont {T.}~\bibnamefont {Claesson}}, \bibinfo {author}
  {\bibfnamefont {O.}~\bibnamefont {Tjernberg}}, \bibinfo {author}
  {\bibfnamefont {A.}~\bibnamefont {Bendounan}}, \bibinfo {author}
  {\bibfnamefont {Y.}~\bibnamefont {Sassa}}, \bibinfo {author} {\bibfnamefont
  {L.}~\bibnamefont {Patthey}}, \bibinfo {author} {\bibfnamefont
  {N.}~\bibnamefont {Momono}}, \bibinfo {author} {\bibfnamefont
  {M.}~\bibnamefont {Oda}}, \bibinfo {author} {\bibfnamefont {M.}~\bibnamefont
  {Ido}}, \bibinfo {author} {\bibfnamefont {S.}~\bibnamefont {Guerrero}},
  \bibinfo {author} {\bibfnamefont {C.}~\bibnamefont {Mudry}}, \ and\ \bibinfo
  {author} {\bibfnamefont {J.}~\bibnamefont {Mesot}},\ }\href {\doibase
  10.1103/PhysRevB.78.205103} {\bibfield  {journal} {\bibinfo  {journal} {Phys.
  Rev. B}\ }\textbf {\bibinfo {volume} {78}},\ \bibinfo {pages} {205103}
  (\bibinfo {year} {2008}{\natexlab{a}})}\BibitemShut {NoStop}%
\bibitem [{\citenamefont {Braicovich}\ \emph {et~al.}(2010)\citenamefont
  {Braicovich}, \citenamefont {van~den Brink}, \citenamefont {Bisogni},
  \citenamefont {Sala}, \citenamefont {Ament}, \citenamefont {Brookes},
  \citenamefont {{De Luca}}, \citenamefont {Salluzzo}, \citenamefont {Schmitt},
  \citenamefont {Strocov},\ and\ \citenamefont
  {Ghiringhelli}}]{BraicovichPRL10}%
  \BibitemOpen
  \bibfield  {author} {\bibinfo {author} {\bibfnamefont {L.}~\bibnamefont
  {Braicovich}}, \bibinfo {author} {\bibfnamefont {J.}~\bibnamefont {van~den
  Brink}}, \bibinfo {author} {\bibfnamefont {V.}~\bibnamefont {Bisogni}},
  \bibinfo {author} {\bibfnamefont {M.~M.}\ \bibnamefont {Sala}}, \bibinfo
  {author} {\bibfnamefont {L.~J.~P.}\ \bibnamefont {Ament}}, \bibinfo {author}
  {\bibfnamefont {N.~B.}\ \bibnamefont {Brookes}}, \bibinfo {author}
  {\bibfnamefont {G.~M.}\ \bibnamefont {{De Luca}}}, \bibinfo {author}
  {\bibfnamefont {M.}~\bibnamefont {Salluzzo}}, \bibinfo {author}
  {\bibfnamefont {T.}~\bibnamefont {Schmitt}}, \bibinfo {author} {\bibfnamefont
  {V.~N.}\ \bibnamefont {Strocov}}, \ and\ \bibinfo {author} {\bibfnamefont
  {G.}~\bibnamefont {Ghiringhelli}},\ }\href {\doibase
  10.1103/PhysRevLett.104.077002} {\bibfield  {journal} {\bibinfo  {journal}
  {Phys. Rev. Lett.}\ }\textbf {\bibinfo {volume} {104}},\ \bibinfo {pages}
  {077002} (\bibinfo {year} {2010})}\BibitemShut {NoStop}%
\bibitem [{\citenamefont {Ghiringhelli}\ \emph {et~al.}(2006)\citenamefont
  {Ghiringhelli}, \citenamefont {Piazzalunga}, \citenamefont {Dallera},
  \citenamefont {Trezzi}, \citenamefont {Braicovich}, \citenamefont {Schmitt},
  \citenamefont {Strocov}, \citenamefont {Betemps}, \citenamefont {Patthey},
  \citenamefont {Wang},\ and\ \citenamefont
  {Grioni}}]{ghiringhelliREVSCIINS2006}%
  \BibitemOpen
  \bibfield  {author} {\bibinfo {author} {\bibfnamefont {G.}~\bibnamefont
  {Ghiringhelli}}, \bibinfo {author} {\bibfnamefont {A.}~\bibnamefont
  {Piazzalunga}}, \bibinfo {author} {\bibfnamefont {C.}~\bibnamefont
  {Dallera}}, \bibinfo {author} {\bibfnamefont {G.}~\bibnamefont {Trezzi}},
  \bibinfo {author} {\bibfnamefont {L.}~\bibnamefont {Braicovich}}, \bibinfo
  {author} {\bibfnamefont {T.}~\bibnamefont {Schmitt}}, \bibinfo {author}
  {\bibfnamefont {V.~N.}\ \bibnamefont {Strocov}}, \bibinfo {author}
  {\bibfnamefont {R.}~\bibnamefont {Betemps}}, \bibinfo {author} {\bibfnamefont
  {L.}~\bibnamefont {Patthey}}, \bibinfo {author} {\bibfnamefont
  {X.}~\bibnamefont {Wang}}, \ and\ \bibinfo {author} {\bibfnamefont
  {M.}~\bibnamefont {Grioni}},\ }\href {\doibase 10.1063/1.2372731} {\bibfield
  {journal} {\bibinfo  {journal} {Review of Scientific Instruments}\ }\textbf
  {\bibinfo {volume} {77}},\ \bibinfo {eid} {113108} (\bibinfo {year}
  {2006})}\BibitemShut {NoStop}%
\bibitem [{\citenamefont {Strocov}\ \emph {et~al.}(2010)\citenamefont
  {Strocov}, \citenamefont {Schmitt}, \citenamefont {Flechsig}, \citenamefont
  {Schmidt}, \citenamefont {Imhof}, \citenamefont {Chen}, \citenamefont
  {Raabe}, \citenamefont {Betemps}, \citenamefont {Zimoch}, \citenamefont
  {Krempasky}, \citenamefont {Wang}, \citenamefont {Grioni},\ and\
  \citenamefont {Patthey}}]{strocovJSYNRAD2010}%
  \BibitemOpen
  \bibfield  {author} {\bibinfo {author} {\bibfnamefont {V.~N.}\ \bibnamefont
  {Strocov}}, \bibinfo {author} {\bibfnamefont {T.}~\bibnamefont {Schmitt}},
  \bibinfo {author} {\bibfnamefont {U.}~\bibnamefont {Flechsig}}, \bibinfo
  {author} {\bibfnamefont {T.}~\bibnamefont {Schmidt}}, \bibinfo {author}
  {\bibfnamefont {A.}~\bibnamefont {Imhof}}, \bibinfo {author} {\bibfnamefont
  {Q.}~\bibnamefont {Chen}}, \bibinfo {author} {\bibfnamefont {J.}~\bibnamefont
  {Raabe}}, \bibinfo {author} {\bibfnamefont {R.}~\bibnamefont {Betemps}},
  \bibinfo {author} {\bibfnamefont {D.}~\bibnamefont {Zimoch}}, \bibinfo
  {author} {\bibfnamefont {J.}~\bibnamefont {Krempasky}}, \bibinfo {author}
  {\bibfnamefont {X.}~\bibnamefont {Wang}}, \bibinfo {author} {\bibfnamefont
  {M.~P.~A.}\ \bibnamefont {Grioni}}, \ and\ \bibinfo {author} {\bibfnamefont
  {L.}~\bibnamefont {Patthey}},\ }\href {\doibase 10.1107/S0909049510019862}
  {\bibfield  {journal} {\bibinfo  {journal} {J. Synchrotron Radiat.}\ }\textbf
  {\bibinfo {volume} {17}},\ \bibinfo {pages} {631} (\bibinfo {year}
  {2010})}\BibitemShut {NoStop}%
\bibitem [{\citenamefont {Komiya}\ \emph {et~al.}(2002)\citenamefont {Komiya},
  \citenamefont {Ando}, \citenamefont {Sun},\ and\ \citenamefont
  {Lavrov}}]{KomiyaPRB02}%
  \BibitemOpen
  \bibfield  {author} {\bibinfo {author} {\bibfnamefont {S.}~\bibnamefont
  {Komiya}}, \bibinfo {author} {\bibfnamefont {Y.}~\bibnamefont {Ando}},
  \bibinfo {author} {\bibfnamefont {X.~F.}\ \bibnamefont {Sun}}, \ and\
  \bibinfo {author} {\bibfnamefont {A.~N.}\ \bibnamefont {Lavrov}},\ }\href
  {\doibase 10.1103/PhysRevB.65.214535} {\bibfield  {journal} {\bibinfo
  {journal} {Phys. Rev. B}\ }\textbf {\bibinfo {volume} {65}},\ \bibinfo
  {pages} {214535} (\bibinfo {year} {2002})}\BibitemShut {NoStop}%
\bibitem [{\citenamefont {Chang}\ \emph {et~al.}(2009)\citenamefont {Chang},
  \citenamefont {Christensen}, \citenamefont {Niedermayer}, \citenamefont
  {Lefmann}, \citenamefont {R{\o}nnow}, \citenamefont {McMorrow}, \citenamefont
  {Schneidewind}, \citenamefont {Link}, \citenamefont {Hiess}, \citenamefont
  {Boehm}, \citenamefont {Mottl}, \citenamefont {Pailh{\'e}s}, \citenamefont
  {Momono}, \citenamefont {Oda}, \citenamefont {Ido},\ and\ \citenamefont
  {Mesot}}]{ChangPRL09}%
  \BibitemOpen
  \bibfield  {author} {\bibinfo {author} {\bibfnamefont {J.}~\bibnamefont
  {Chang}}, \bibinfo {author} {\bibfnamefont {N.~B.}\ \bibnamefont
  {Christensen}}, \bibinfo {author} {\bibfnamefont {C.}~\bibnamefont
  {Niedermayer}}, \bibinfo {author} {\bibfnamefont {K.}~\bibnamefont
  {Lefmann}}, \bibinfo {author} {\bibfnamefont {H.~M.}\ \bibnamefont
  {R{\o}nnow}}, \bibinfo {author} {\bibfnamefont {D.~F.}\ \bibnamefont
  {McMorrow}}, \bibinfo {author} {\bibfnamefont {A.}~\bibnamefont
  {Schneidewind}}, \bibinfo {author} {\bibfnamefont {P.}~\bibnamefont {Link}},
  \bibinfo {author} {\bibfnamefont {A.}~\bibnamefont {Hiess}}, \bibinfo
  {author} {\bibfnamefont {M.}~\bibnamefont {Boehm}}, \bibinfo {author}
  {\bibfnamefont {R.}~\bibnamefont {Mottl}}, \bibinfo {author} {\bibfnamefont
  {S.}~\bibnamefont {Pailh{\'e}s}}, \bibinfo {author} {\bibfnamefont
  {N.}~\bibnamefont {Momono}}, \bibinfo {author} {\bibfnamefont
  {M.}~\bibnamefont {Oda}}, \bibinfo {author} {\bibfnamefont {M.}~\bibnamefont
  {Ido}}, \ and\ \bibinfo {author} {\bibfnamefont {J.}~\bibnamefont {Mesot}},\
  }\href {\doibase 10.1103/PhysRevLett.102.177006} {\bibfield  {journal}
  {\bibinfo  {journal} {Phys. Rev. Lett.}\ }\textbf {\bibinfo {volume} {102}},\
  \bibinfo {pages} {177006} (\bibinfo {year} {2009})}\BibitemShut {NoStop}%
\bibitem [{\citenamefont {R{\o}mer}\ \emph {et~al.}(2013)\citenamefont
  {R{\o}mer}, \citenamefont {Chang}, \citenamefont {Christensen}, \citenamefont
  {Andersen}, \citenamefont {Lefmann}, \citenamefont {M{\"a}hler},
  \citenamefont {Gavilano}, \citenamefont {Gilardi}, \citenamefont
  {Niedermayer}, \citenamefont {R{\o}nnow}, \citenamefont {Schneidewind},
  \citenamefont {Link}, \citenamefont {Oda}, \citenamefont {Ido}, \citenamefont
  {Momono},\ and\ \citenamefont {Mesot}}]{RomerPRB13}%
  \BibitemOpen
  \bibfield  {author} {\bibinfo {author} {\bibfnamefont {A.~T.}\ \bibnamefont
  {R{\o}mer}}, \bibinfo {author} {\bibfnamefont {J.}~\bibnamefont {Chang}},
  \bibinfo {author} {\bibfnamefont {N.~B.}\ \bibnamefont {Christensen}},
  \bibinfo {author} {\bibfnamefont {B.~M.}\ \bibnamefont {Andersen}}, \bibinfo
  {author} {\bibfnamefont {K.}~\bibnamefont {Lefmann}}, \bibinfo {author}
  {\bibfnamefont {L.}~\bibnamefont {M{\"a}hler}}, \bibinfo {author}
  {\bibfnamefont {J.}~\bibnamefont {Gavilano}}, \bibinfo {author}
  {\bibfnamefont {R.}~\bibnamefont {Gilardi}}, \bibinfo {author} {\bibfnamefont
  {C.}~\bibnamefont {Niedermayer}}, \bibinfo {author} {\bibfnamefont {H.~M.}\
  \bibnamefont {R{\o}nnow}}, \bibinfo {author} {\bibfnamefont {A.}~\bibnamefont
  {Schneidewind}}, \bibinfo {author} {\bibfnamefont {P.}~\bibnamefont {Link}},
  \bibinfo {author} {\bibfnamefont {M.}~\bibnamefont {Oda}}, \bibinfo {author}
  {\bibfnamefont {M.}~\bibnamefont {Ido}}, \bibinfo {author} {\bibfnamefont
  {N.}~\bibnamefont {Momono}}, \ and\ \bibinfo {author} {\bibfnamefont
  {J.}~\bibnamefont {Mesot}},\ }\href {\doibase 10.1103/PhysRevB.87.144513}
  {\bibfield  {journal} {\bibinfo  {journal} {Phys. Rev. B}\ }\textbf {\bibinfo
  {volume} {87}},\ \bibinfo {pages} {144513} (\bibinfo {year}
  {2013})}\BibitemShut {NoStop}%
\bibitem [{\citenamefont {Chang}\ \emph {et~al.}(2012)\citenamefont {Chang},
  \citenamefont {White}, \citenamefont {Laver}, \citenamefont {Bowell},
  \citenamefont {Brown}, \citenamefont {Holmes}, \citenamefont {Maechler},
  \citenamefont {Str{\"a}ssle}, \citenamefont {Gilardi}, \citenamefont
  {Gerber}, \citenamefont {Kurosawa}, \citenamefont {Momono}, \citenamefont
  {Oda}, \citenamefont {Ido}, \citenamefont {Lipscombe}, \citenamefont
  {Hayden}, \citenamefont {Dewhurst}, \citenamefont {Vavrin}, \citenamefont
  {Gavilano}, \citenamefont {Kohlbrecher}, \citenamefont {Forgan},\ and\
  \citenamefont {Mesot}}]{ChangPRB12}%
  \BibitemOpen
  \bibfield  {author} {\bibinfo {author} {\bibfnamefont {J.}~\bibnamefont
  {Chang}}, \bibinfo {author} {\bibfnamefont {J.~S.}\ \bibnamefont {White}},
  \bibinfo {author} {\bibfnamefont {M.}~\bibnamefont {Laver}}, \bibinfo
  {author} {\bibfnamefont {C.~J.}\ \bibnamefont {Bowell}}, \bibinfo {author}
  {\bibfnamefont {S.~P.}\ \bibnamefont {Brown}}, \bibinfo {author}
  {\bibfnamefont {A.~T.}\ \bibnamefont {Holmes}}, \bibinfo {author}
  {\bibfnamefont {L.}~\bibnamefont {Maechler}}, \bibinfo {author}
  {\bibfnamefont {S.}~\bibnamefont {Str{\"a}ssle}}, \bibinfo {author}
  {\bibfnamefont {R.}~\bibnamefont {Gilardi}}, \bibinfo {author} {\bibfnamefont
  {S.}~\bibnamefont {Gerber}}, \bibinfo {author} {\bibfnamefont
  {T.}~\bibnamefont {Kurosawa}}, \bibinfo {author} {\bibfnamefont
  {N.}~\bibnamefont {Momono}}, \bibinfo {author} {\bibfnamefont
  {M.}~\bibnamefont {Oda}}, \bibinfo {author} {\bibfnamefont {M.}~\bibnamefont
  {Ido}}, \bibinfo {author} {\bibfnamefont {O.~J.}\ \bibnamefont {Lipscombe}},
  \bibinfo {author} {\bibfnamefont {S.~M.}\ \bibnamefont {Hayden}}, \bibinfo
  {author} {\bibfnamefont {C.~D.}\ \bibnamefont {Dewhurst}}, \bibinfo {author}
  {\bibfnamefont {R.}~\bibnamefont {Vavrin}}, \bibinfo {author} {\bibfnamefont
  {J.}~\bibnamefont {Gavilano}}, \bibinfo {author} {\bibfnamefont
  {J.}~\bibnamefont {Kohlbrecher}}, \bibinfo {author} {\bibfnamefont {E.~M.}\
  \bibnamefont {Forgan}}, \ and\ \bibinfo {author} {\bibfnamefont
  {J.}~\bibnamefont {Mesot}},\ }\href {\doibase 10.1103/PhysRevB.85.134520}
  {\bibfield  {journal} {\bibinfo  {journal} {Phys. Rev. B}\ }\textbf {\bibinfo
  {volume} {85}},\ \bibinfo {pages} {134520} (\bibinfo {year}
  {2012})}\BibitemShut {NoStop}%
\bibitem [{\citenamefont {Chang}\ \emph
  {et~al.}(2008{\natexlab{b}})\citenamefont {Chang}, \citenamefont
  {Niedermayer}, \citenamefont {Gilardi}, \citenamefont {Christensen},
  \citenamefont {R{\o}nnow}, \citenamefont {McMorrow}, \citenamefont {Ay},
  \citenamefont {Stahn}, \citenamefont {Sobolev}, \citenamefont {Hiess},
  \citenamefont {Pailhes}, \citenamefont {Baines}, \citenamefont {Momono},
  \citenamefont {Oda}, \citenamefont {Ido},\ and\ \citenamefont
  {Mesot}}]{ChangPRB08}%
  \BibitemOpen
  \bibfield  {author} {\bibinfo {author} {\bibfnamefont {J.}~\bibnamefont
  {Chang}}, \bibinfo {author} {\bibfnamefont {C.}~\bibnamefont {Niedermayer}},
  \bibinfo {author} {\bibfnamefont {R.}~\bibnamefont {Gilardi}}, \bibinfo
  {author} {\bibfnamefont {N.~B.}\ \bibnamefont {Christensen}}, \bibinfo
  {author} {\bibfnamefont {H.~M.}\ \bibnamefont {R{\o}nnow}}, \bibinfo {author}
  {\bibfnamefont {D.~F.}\ \bibnamefont {McMorrow}}, \bibinfo {author}
  {\bibfnamefont {M.}~\bibnamefont {Ay}}, \bibinfo {author} {\bibfnamefont
  {J.}~\bibnamefont {Stahn}}, \bibinfo {author} {\bibfnamefont
  {O.}~\bibnamefont {Sobolev}}, \bibinfo {author} {\bibfnamefont
  {A.}~\bibnamefont {Hiess}}, \bibinfo {author} {\bibfnamefont
  {S.}~\bibnamefont {Pailhes}}, \bibinfo {author} {\bibfnamefont
  {C.}~\bibnamefont {Baines}}, \bibinfo {author} {\bibfnamefont
  {N.}~\bibnamefont {Momono}}, \bibinfo {author} {\bibfnamefont
  {M.}~\bibnamefont {Oda}}, \bibinfo {author} {\bibfnamefont {M.}~\bibnamefont
  {Ido}}, \ and\ \bibinfo {author} {\bibfnamefont {J.}~\bibnamefont {Mesot}},\
  }\href {\doibase 10.1103/PhysRevB.78.104525} {\bibfield  {journal} {\bibinfo
  {journal} {Phys. Rev. B}\ }\textbf {\bibinfo {volume} {78}},\ \bibinfo
  {pages} {104525} (\bibinfo {year} {2008}{\natexlab{b}})}\BibitemShut
  {NoStop}%
\bibitem [{\citenamefont {Thampy}\ \emph {et~al.}(2014)\citenamefont {Thampy},
  \citenamefont {Dean}, \citenamefont {Christensen}, \citenamefont {Steinke},
  \citenamefont {Islam}, \citenamefont {Oda}, \citenamefont {Ido},
  \citenamefont {Momono}, \citenamefont {Wilkins},\ and\ \citenamefont
  {Hill}}]{ThampyPRB2014}%
  \BibitemOpen
  \bibfield  {author} {\bibinfo {author} {\bibfnamefont {V.}~\bibnamefont
  {Thampy}}, \bibinfo {author} {\bibfnamefont {M.~P.~M.}\ \bibnamefont {Dean}},
  \bibinfo {author} {\bibfnamefont {N.~B.}\ \bibnamefont {Christensen}},
  \bibinfo {author} {\bibfnamefont {L.}~\bibnamefont {Steinke}}, \bibinfo
  {author} {\bibfnamefont {Z.}~\bibnamefont {Islam}}, \bibinfo {author}
  {\bibfnamefont {M.}~\bibnamefont {Oda}}, \bibinfo {author} {\bibfnamefont
  {M.}~\bibnamefont {Ido}}, \bibinfo {author} {\bibfnamefont {N.}~\bibnamefont
  {Momono}}, \bibinfo {author} {\bibfnamefont {S.~B.}\ \bibnamefont {Wilkins}},
  \ and\ \bibinfo {author} {\bibfnamefont {J.~P.}\ \bibnamefont {Hill}},\
  }\href {\doibase 10.1103/PhysRevB.90.100510} {\bibfield  {journal} {\bibinfo
  {journal} {Phys. Rev. B}\ }\textbf {\bibinfo {volume} {90}},\ \bibinfo
  {pages} {100510} (\bibinfo {year} {2014})}\BibitemShut {NoStop}%
\bibitem [{\citenamefont {Croft}\ \emph {et~al.}(2014)\citenamefont {Croft},
  \citenamefont {Lester}, \citenamefont {Senn}, \citenamefont {Bombardi},\ and\
  \citenamefont {Hayden}}]{HaydenPRB2014}%
  \BibitemOpen
  \bibfield  {author} {\bibinfo {author} {\bibfnamefont {T.~P.}\ \bibnamefont
  {Croft}}, \bibinfo {author} {\bibfnamefont {C.}~\bibnamefont {Lester}},
  \bibinfo {author} {\bibfnamefont {M.~S.}\ \bibnamefont {Senn}}, \bibinfo
  {author} {\bibfnamefont {A.}~\bibnamefont {Bombardi}}, \ and\ \bibinfo
  {author} {\bibfnamefont {S.~M.}\ \bibnamefont {Hayden}},\ }\href {\doibase
  10.1103/PhysRevB.89.224513} {\bibfield  {journal} {\bibinfo  {journal} {Phys.
  Rev. B}\ }\textbf {\bibinfo {volume} {89}},\ \bibinfo {pages} {224513}
  (\bibinfo {year} {2014})}\BibitemShut {NoStop}%
\bibitem [{\citenamefont {Monney}\ \emph {et~al.}(2016)\citenamefont {Monney},
  \citenamefont {Schmitt}, \citenamefont {Matt}, \citenamefont {Mesot},
  \citenamefont {Strocov}, \citenamefont {Lipscombe}, \citenamefont {Hayden},\
  and\ \citenamefont {Chang}}]{MonneyPRB16}%
  \BibitemOpen
  \bibfield  {author} {\bibinfo {author} {\bibfnamefont {C.}~\bibnamefont
  {Monney}}, \bibinfo {author} {\bibfnamefont {T.}~\bibnamefont {Schmitt}},
  \bibinfo {author} {\bibfnamefont {C.~E.}\ \bibnamefont {Matt}}, \bibinfo
  {author} {\bibfnamefont {J.}~\bibnamefont {Mesot}}, \bibinfo {author}
  {\bibfnamefont {V.~N.}\ \bibnamefont {Strocov}}, \bibinfo {author}
  {\bibfnamefont {O.~J.}\ \bibnamefont {Lipscombe}}, \bibinfo {author}
  {\bibfnamefont {S.~M.}\ \bibnamefont {Hayden}}, \ and\ \bibinfo {author}
  {\bibfnamefont {J.}~\bibnamefont {Chang}},\ }\href {\doibase
  10.1103/PhysRevB.93.075103} {\bibfield  {journal} {\bibinfo  {journal} {Phys.
  Rev. B}\ }\textbf {\bibinfo {volume} {93}},\ \bibinfo {pages} {075103}
  (\bibinfo {year} {2016})}\BibitemShut {NoStop}%
\bibitem [{\citenamefont {Lamsal}\ and\ \citenamefont
  {Montfrooij}(2016)}]{LamsalPRB16}%
  \BibitemOpen
  \bibfield  {author} {\bibinfo {author} {\bibfnamefont {J.}~\bibnamefont
  {Lamsal}}\ and\ \bibinfo {author} {\bibfnamefont {W.}~\bibnamefont
  {Montfrooij}},\ }\href {\doibase 10.1103/PhysRevB.93.214513} {\bibfield
  {journal} {\bibinfo  {journal} {Phys. Rev. B}\ }\textbf {\bibinfo {volume}
  {93}},\ \bibinfo {pages} {214513} (\bibinfo {year} {2016})}\BibitemShut
  {NoStop}%
\bibitem [{\citenamefont {{Miao}}\ \emph {et~al.}(2017)\citenamefont {{Miao}},
  \citenamefont {{Lorenzana}}, \citenamefont {{Seibold}}, \citenamefont
  {{Peng}}, \citenamefont {{Amorese}}, \citenamefont {{Yakhou-Harris}},
  \citenamefont {{Kummer}}, \citenamefont {{Brookes}}, \citenamefont {{Konik}},
  \citenamefont {{Thampy}}, \citenamefont {{Gu}}, \citenamefont
  {{Ghiringhelli}}, \citenamefont {{Braicovich}},\ and\ \citenamefont
  {{Dean}}}]{Miao17}%
  \BibitemOpen
  \bibfield  {author} {\bibinfo {author} {\bibfnamefont {H.}~\bibnamefont
  {{Miao}}}, \bibinfo {author} {\bibfnamefont {J.}~\bibnamefont {{Lorenzana}}},
  \bibinfo {author} {\bibfnamefont {G.}~\bibnamefont {{Seibold}}}, \bibinfo
  {author} {\bibfnamefont {Y.~Y.}\ \bibnamefont {{Peng}}}, \bibinfo {author}
  {\bibfnamefont {A.}~\bibnamefont {{Amorese}}}, \bibinfo {author}
  {\bibfnamefont {F.}~\bibnamefont {{Yakhou-Harris}}}, \bibinfo {author}
  {\bibfnamefont {K.}~\bibnamefont {{Kummer}}}, \bibinfo {author}
  {\bibfnamefont {N.~B.}\ \bibnamefont {{Brookes}}}, \bibinfo {author}
  {\bibfnamefont {R.~M.}\ \bibnamefont {{Konik}}}, \bibinfo {author}
  {\bibfnamefont {V.}~\bibnamefont {{Thampy}}}, \bibinfo {author}
  {\bibfnamefont {G.~D.}\ \bibnamefont {{Gu}}}, \bibinfo {author}
  {\bibfnamefont {G.}~\bibnamefont {{Ghiringhelli}}}, \bibinfo {author}
  {\bibfnamefont {L.}~\bibnamefont {{Braicovich}}}, \ and\ \bibinfo {author}
  {\bibfnamefont {M.~P.~M.}\ \bibnamefont {{Dean}}},\ }\href@noop {} {\bibfield
   {journal} {\bibinfo  {journal} {ArXiv e-prints}\ } (\bibinfo {year}
  {2017})},\ \Eprint {http://arxiv.org/abs/1701.00022} {arXiv:1701.00022
  [cond-mat.supr-con]} \BibitemShut {NoStop}%
\bibitem [{\citenamefont {Braicovich}\ \emph {et~al.}(2009)\citenamefont
  {Braicovich}, \citenamefont {Ament}, \citenamefont {Bisogni}, \citenamefont
  {Forte}, \citenamefont {Aruta}, \citenamefont {Balestrino}, \citenamefont
  {Brookes}, \citenamefont {{De Luca}}, \citenamefont {Medaglia}, \citenamefont
  {Granozio}, \citenamefont {Radovic}, \citenamefont {Salluzzo}, \citenamefont
  {van~den Brink},\ and\ \citenamefont {Ghiringhelli}}]{BraicovichPRL09}%
  \BibitemOpen
  \bibfield  {author} {\bibinfo {author} {\bibfnamefont {L.}~\bibnamefont
  {Braicovich}}, \bibinfo {author} {\bibfnamefont {L.~J.~P.}\ \bibnamefont
  {Ament}}, \bibinfo {author} {\bibfnamefont {V.}~\bibnamefont {Bisogni}},
  \bibinfo {author} {\bibfnamefont {F.}~\bibnamefont {Forte}}, \bibinfo
  {author} {\bibfnamefont {C.}~\bibnamefont {Aruta}}, \bibinfo {author}
  {\bibfnamefont {G.}~\bibnamefont {Balestrino}}, \bibinfo {author}
  {\bibfnamefont {N.~B.}\ \bibnamefont {Brookes}}, \bibinfo {author}
  {\bibfnamefont {G.~M.}\ \bibnamefont {{De Luca}}}, \bibinfo {author}
  {\bibfnamefont {P.~G.}\ \bibnamefont {Medaglia}}, \bibinfo {author}
  {\bibfnamefont {F.~M.}\ \bibnamefont {Granozio}}, \bibinfo {author}
  {\bibfnamefont {M.}~\bibnamefont {Radovic}}, \bibinfo {author} {\bibfnamefont
  {M.}~\bibnamefont {Salluzzo}}, \bibinfo {author} {\bibfnamefont
  {J.}~\bibnamefont {van~den Brink}}, \ and\ \bibinfo {author} {\bibfnamefont
  {G.}~\bibnamefont {Ghiringhelli}},\ }\href {\doibase
  10.1103/PhysRevLett.102.167401} {\bibfield  {journal} {\bibinfo  {journal}
  {Phys. Rev. Lett.}\ }\textbf {\bibinfo {volume} {102}},\ \bibinfo {pages}
  {167401} (\bibinfo {year} {2009})}\BibitemShut {NoStop}%
\bibitem [{\citenamefont {Guarise}\ \emph {et~al.}(2014)\citenamefont
  {Guarise}, \citenamefont {Piazza}, \citenamefont {Berger}, \citenamefont
  {Giannini}, \citenamefont {Schmitt}, \citenamefont {R{\o}nnow}, \citenamefont
  {Sawatzky}, \citenamefont {van~den Brink}, \citenamefont {Altenfeld},
  \citenamefont {Eremin},\ and\ \citenamefont {Grioni}}]{GuariseNATC2014}%
  \BibitemOpen
  \bibfield  {author} {\bibinfo {author} {\bibfnamefont {M.}~\bibnamefont
  {Guarise}}, \bibinfo {author} {\bibfnamefont {B.~D.}\ \bibnamefont {Piazza}},
  \bibinfo {author} {\bibfnamefont {H.}~\bibnamefont {Berger}}, \bibinfo
  {author} {\bibfnamefont {E.}~\bibnamefont {Giannini}}, \bibinfo {author}
  {\bibfnamefont {T.}~\bibnamefont {Schmitt}}, \bibinfo {author} {\bibfnamefont
  {H.~M.}\ \bibnamefont {R{\o}nnow}}, \bibinfo {author} {\bibfnamefont {G.~A.}\
  \bibnamefont {Sawatzky}}, \bibinfo {author} {\bibfnamefont {J.}~\bibnamefont
  {van~den Brink}}, \bibinfo {author} {\bibfnamefont {D.}~\bibnamefont
  {Altenfeld}}, \bibinfo {author} {\bibfnamefont {I.}~\bibnamefont {Eremin}}, \
  and\ \bibinfo {author} {\bibfnamefont {M.}~\bibnamefont {Grioni}},\ }\href
  {\doibase 10.1038/ncomms6760;;;;;;;;;;;;;;;;;;;;;;;;;;;;;;;;;
  10.1038/ncomms6760} {\bibfield  {journal} {\bibinfo  {journal} {Nat Commun}\
  }\textbf {\bibinfo {volume} {5}},\ \bibinfo {pages} {5760} (\bibinfo {year}
  {2014})}\BibitemShut {NoStop}%
\bibitem [{\citenamefont {Dean}\ \emph {et~al.}(2014)\citenamefont {Dean},
  \citenamefont {James}, \citenamefont {Walters}, \citenamefont {Bisogni},
  \citenamefont {Jarrige}, \citenamefont {H{\"u}cker}, \citenamefont
  {Giannini}, \citenamefont {Fujita}, \citenamefont {Pelliciari}, \citenamefont
  {Huang}, \citenamefont {Konik}, \citenamefont {Schmitt},\ and\ \citenamefont
  {Hill}}]{DeanPRB2014}%
  \BibitemOpen
  \bibfield  {author} {\bibinfo {author} {\bibfnamefont {M.~P.~M.}\
  \bibnamefont {Dean}}, \bibinfo {author} {\bibfnamefont {A.~J.~A.}\
  \bibnamefont {James}}, \bibinfo {author} {\bibfnamefont {A.~C.}\ \bibnamefont
  {Walters}}, \bibinfo {author} {\bibfnamefont {V.}~\bibnamefont {Bisogni}},
  \bibinfo {author} {\bibfnamefont {I.}~\bibnamefont {Jarrige}}, \bibinfo
  {author} {\bibfnamefont {M.}~\bibnamefont {H{\"u}cker}}, \bibinfo {author}
  {\bibfnamefont {E.}~\bibnamefont {Giannini}}, \bibinfo {author}
  {\bibfnamefont {M.}~\bibnamefont {Fujita}}, \bibinfo {author} {\bibfnamefont
  {J.}~\bibnamefont {Pelliciari}}, \bibinfo {author} {\bibfnamefont {Y.~B.}\
  \bibnamefont {Huang}}, \bibinfo {author} {\bibfnamefont {R.~M.}\ \bibnamefont
  {Konik}}, \bibinfo {author} {\bibfnamefont {T.}~\bibnamefont {Schmitt}}, \
  and\ \bibinfo {author} {\bibfnamefont {J.~P.}\ \bibnamefont {Hill}},\ }\href
  {\doibase 10.1103/PhysRevB.90.220506} {\bibfield  {journal} {\bibinfo
  {journal} {Phys. Rev. B}\ }\textbf {\bibinfo {volume} {90}},\ \bibinfo
  {pages} {220506} (\bibinfo {year} {2014})}\BibitemShut {NoStop}%
\bibitem [{\citenamefont {Peng}\ \emph {et~al.}(2016)\citenamefont {Peng},
  \citenamefont {Dellea}, \citenamefont {Minola}, \citenamefont {Conni},
  \citenamefont {Amorese}, \citenamefont {{Di Castro}}, \citenamefont {{De
  Luca}}, \citenamefont {Kummer}, \citenamefont {Salluzzo}, \citenamefont
  {Sun}, \citenamefont {Zhou}, \citenamefont {Balestrino}, \citenamefont
  {Tacon}, \citenamefont {Keimer}, \citenamefont {Braicovich}, \citenamefont
  {Brookes},\ and\ \citenamefont {Ghiringhelli}}]{Peng16}%
  \BibitemOpen
  \bibfield  {author} {\bibinfo {author} {\bibfnamefont {Y.~Y.}\ \bibnamefont
  {Peng}}, \bibinfo {author} {\bibfnamefont {G.}~\bibnamefont {Dellea}},
  \bibinfo {author} {\bibfnamefont {M.}~\bibnamefont {Minola}}, \bibinfo
  {author} {\bibfnamefont {M.}~\bibnamefont {Conni}}, \bibinfo {author}
  {\bibfnamefont {A.}~\bibnamefont {Amorese}}, \bibinfo {author} {\bibfnamefont
  {D.}~\bibnamefont {{Di Castro}}}, \bibinfo {author} {\bibfnamefont {G.~M.}\
  \bibnamefont {{De Luca}}}, \bibinfo {author} {\bibfnamefont {K.}~\bibnamefont
  {Kummer}}, \bibinfo {author} {\bibfnamefont {M.}~\bibnamefont {Salluzzo}},
  \bibinfo {author} {\bibfnamefont {X.}~\bibnamefont {Sun}}, \bibinfo {author}
  {\bibfnamefont {X.~J.}\ \bibnamefont {Zhou}}, \bibinfo {author}
  {\bibfnamefont {G.}~\bibnamefont {Balestrino}}, \bibinfo {author}
  {\bibfnamefont {M.~L.}\ \bibnamefont {Tacon}}, \bibinfo {author}
  {\bibfnamefont {B.}~\bibnamefont {Keimer}}, \bibinfo {author} {\bibfnamefont
  {L.}~\bibnamefont {Braicovich}}, \bibinfo {author} {\bibfnamefont {N.~B.}\
  \bibnamefont {Brookes}}, \ and\ \bibinfo {author} {\bibfnamefont
  {G.}~\bibnamefont {Ghiringhelli}},\ }\href {http://arxiv.org/abs/1609.05405}
  {\bibfield  {journal} {\bibinfo  {journal} {arXiv}\ } (\bibinfo {year}
  {2016})},\ \Eprint {http://arxiv.org/abs/1609.05405} {1609.05405}
  \BibitemShut {NoStop}%
\bibitem [{\citenamefont {{Hozoi Liviu}}\ \emph {et~al.}(2011)\citenamefont
  {{Hozoi Liviu}}, \citenamefont {{Siurakshina Liudmila}}, \citenamefont
  {{Fulde Peter}},\ and\ \citenamefont {Jeroen}}]{HozoiSREP11}%
  \BibitemOpen
  \bibfield  {author} {\bibinfo {author} {\bibnamefont {{Hozoi Liviu}}},
  \bibinfo {author} {\bibnamefont {{Siurakshina Liudmila}}}, \bibinfo {author}
  {\bibnamefont {{Fulde Peter}}}, \ and\ \bibinfo {author} {\bibfnamefont
  {v.~d.~B.}\ \bibnamefont {Jeroen}},\ }\href {\doibase 10.1038/srep00065;;;;;
  10.1038/srep00065} {\bibfield  {journal} {\bibinfo  {journal} {Scientific
  Reports}\ }\textbf {\bibinfo {volume} {1}},\ \bibinfo {pages} {65} (\bibinfo
  {year} {2011})}\BibitemShut {NoStop}%
\bibitem [{\citenamefont {Sala}\ \emph {et~al.}(2011)\citenamefont {Sala},
  \citenamefont {Bisogni}, \citenamefont {Aruta}, \citenamefont {Balestrino},
  \citenamefont {Berger}, \citenamefont {Brookes}, \citenamefont {de~Luca},
  \citenamefont {Castro}, \citenamefont {Grioni}, \citenamefont {Guarise},
  \citenamefont {Medaglia}, \citenamefont {Granozio}, \citenamefont {Minola},
  \citenamefont {Perna}, \citenamefont {Radovic}, \citenamefont {Salluzzo},
  \citenamefont {Schmitt}, \citenamefont {Zhou}, \citenamefont {Braicovich},\
  and\ \citenamefont {Ghiringhelli}}]{MorettiNJP11}%
  \BibitemOpen
  \bibfield  {author} {\bibinfo {author} {\bibfnamefont {M.~M.}\ \bibnamefont
  {Sala}}, \bibinfo {author} {\bibfnamefont {V.}~\bibnamefont {Bisogni}},
  \bibinfo {author} {\bibfnamefont {C.}~\bibnamefont {Aruta}}, \bibinfo
  {author} {\bibfnamefont {G.}~\bibnamefont {Balestrino}}, \bibinfo {author}
  {\bibfnamefont {H.}~\bibnamefont {Berger}}, \bibinfo {author} {\bibfnamefont
  {N.~B.}\ \bibnamefont {Brookes}}, \bibinfo {author} {\bibfnamefont {G.~M.}\
  \bibnamefont {de~Luca}}, \bibinfo {author} {\bibfnamefont {D.~D.}\
  \bibnamefont {Castro}}, \bibinfo {author} {\bibfnamefont {M.}~\bibnamefont
  {Grioni}}, \bibinfo {author} {\bibfnamefont {M.}~\bibnamefont {Guarise}},
  \bibinfo {author} {\bibfnamefont {P.~G.}\ \bibnamefont {Medaglia}}, \bibinfo
  {author} {\bibfnamefont {F.~M.}\ \bibnamefont {Granozio}}, \bibinfo {author}
  {\bibfnamefont {M.}~\bibnamefont {Minola}}, \bibinfo {author} {\bibfnamefont
  {P.}~\bibnamefont {Perna}}, \bibinfo {author} {\bibfnamefont
  {M.}~\bibnamefont {Radovic}}, \bibinfo {author} {\bibfnamefont
  {M.}~\bibnamefont {Salluzzo}}, \bibinfo {author} {\bibfnamefont
  {T.}~\bibnamefont {Schmitt}}, \bibinfo {author} {\bibfnamefont {K.~J.}\
  \bibnamefont {Zhou}}, \bibinfo {author} {\bibfnamefont {L.}~\bibnamefont
  {Braicovich}}, \ and\ \bibinfo {author} {\bibfnamefont {G.}~\bibnamefont
  {Ghiringhelli}},\ }\href {http://stacks.iop.org/1367-2630/13/i=4/a=043026}
  {\bibfield  {journal} {\bibinfo  {journal} {New Journal of Physics}\ }\textbf
  {\bibinfo {volume} {13}},\ \bibinfo {pages} {043026} (\bibinfo {year}
  {2011})}\BibitemShut {NoStop}%
\bibitem [{\citenamefont {{C. J. Jia}}\ \emph {et~al.}(2014)\citenamefont {{C.
  J. Jia}}, \citenamefont {{E. A. Nowadnick}}, \citenamefont {{K. Wohlfeld}},
  \citenamefont {{Y. F. Kung}}, \citenamefont {{C.-C. Chen}}, \citenamefont
  {{S. Johnston}}, \citenamefont {{T. Tohyama}}, \citenamefont {{B. Moritz}},\
  and\ \citenamefont {{T. P. Devereaux}}}]{JiaNATC2014}%
  \BibitemOpen
  \bibfield  {author} {\bibinfo {author} {\bibnamefont {{C. J. Jia}}}, \bibinfo
  {author} {\bibnamefont {{E. A. Nowadnick}}}, \bibinfo {author} {\bibnamefont
  {{K. Wohlfeld}}}, \bibinfo {author} {\bibnamefont {{Y. F. Kung}}}, \bibinfo
  {author} {\bibnamefont {{C.-C. Chen}}}, \bibinfo {author} {\bibnamefont {{S.
  Johnston}}}, \bibinfo {author} {\bibnamefont {{T. Tohyama}}}, \bibinfo
  {author} {\bibnamefont {{B. Moritz}}}, \ and\ \bibinfo {author} {\bibnamefont
  {{T. P. Devereaux}}},\ }\href {\doibase 10.1038/ncomms4314;;;;;;;;;;;
  10.1038/ncomms4314} {\bibfield  {journal} {\bibinfo  {journal} {Nature
  Communications}\ }\textbf {\bibinfo {volume} {5}},\ \bibinfo {pages} {3314}
  (\bibinfo {year} {2014})}\BibitemShut {NoStop}%
\bibitem [{\citenamefont {Fatuzzo}\ \emph {et~al.}(2014)\citenamefont
  {Fatuzzo}, \citenamefont {Sassa}, \citenamefont {M{\aa}nsson}, \citenamefont
  {Pailh{\`e}s}, \citenamefont {Lipscombe}, \citenamefont {Hayden},
  \citenamefont {Patthey}, \citenamefont {Shi}, \citenamefont {Grioni},
  \citenamefont {R{\o}nnow}, \citenamefont {Mesot}, \citenamefont {Tjernberg},\
  and\ \citenamefont {Chang}}]{FatuzzoPRB2014}%
  \BibitemOpen
  \bibfield  {author} {\bibinfo {author} {\bibfnamefont {C.~G.}\ \bibnamefont
  {Fatuzzo}}, \bibinfo {author} {\bibfnamefont {Y.}~\bibnamefont {Sassa}},
  \bibinfo {author} {\bibfnamefont {M.}~\bibnamefont {M{\aa}nsson}}, \bibinfo
  {author} {\bibfnamefont {S.}~\bibnamefont {Pailh{\`e}s}}, \bibinfo {author}
  {\bibfnamefont {O.~J.}\ \bibnamefont {Lipscombe}}, \bibinfo {author}
  {\bibfnamefont {S.~M.}\ \bibnamefont {Hayden}}, \bibinfo {author}
  {\bibfnamefont {L.}~\bibnamefont {Patthey}}, \bibinfo {author} {\bibfnamefont
  {M.}~\bibnamefont {Shi}}, \bibinfo {author} {\bibfnamefont {M.}~\bibnamefont
  {Grioni}}, \bibinfo {author} {\bibfnamefont {H.~M.}\ \bibnamefont
  {R{\o}nnow}}, \bibinfo {author} {\bibfnamefont {J.}~\bibnamefont {Mesot}},
  \bibinfo {author} {\bibfnamefont {O.}~\bibnamefont {Tjernberg}}, \ and\
  \bibinfo {author} {\bibfnamefont {J.}~\bibnamefont {Chang}},\ }\href
  {\doibase 10.1103/PhysRevB.89.205104} {\bibfield  {journal} {\bibinfo
  {journal} {Phys. Rev. B}\ }\textbf {\bibinfo {volume} {89}},\ \bibinfo
  {pages} {205104} (\bibinfo {year} {2014})}\BibitemShut {NoStop}%
\bibitem [{\citenamefont {{J. Chang}}\ \emph {et~al.}(2013)\citenamefont {{J.
  Chang}}, \citenamefont {{M. M{\aa}nsson}}, \citenamefont {{S. Pailh{\`e}s}},
  \citenamefont {{T. Claesson}}, \citenamefont {{O. J. Lipscombe}},
  \citenamefont {{S. M. Hayden}}, \citenamefont {{L. Patthey}}, \citenamefont
  {{O. Tjernberg}},\ and\ \citenamefont {{J. Mesot}}}]{ChangNATC13}%
  \BibitemOpen
  \bibfield  {author} {\bibinfo {author} {\bibnamefont {{J. Chang}}}, \bibinfo
  {author} {\bibnamefont {{M. M{\aa}nsson}}}, \bibinfo {author} {\bibnamefont
  {{S. Pailh{\`e}s}}}, \bibinfo {author} {\bibnamefont {{T. Claesson}}},
  \bibinfo {author} {\bibnamefont {{O. J. Lipscombe}}}, \bibinfo {author}
  {\bibnamefont {{S. M. Hayden}}}, \bibinfo {author} {\bibnamefont {{L.
  Patthey}}}, \bibinfo {author} {\bibnamefont {{O. Tjernberg}}}, \ and\
  \bibinfo {author} {\bibnamefont {{J. Mesot}}},\ }\href {\doibase
  10.1038/ncomms3559} {\bibfield  {journal} {\bibinfo  {journal} {Nature
  Communications}\ }\textbf {\bibinfo {volume} {4}},\ \bibinfo {pages} {2559}
  (\bibinfo {year} {2013})}\BibitemShut {NoStop}%
\bibitem [{\citenamefont {Plat{\'e}}\ \emph {et~al.}(2005)\citenamefont
  {Plat{\'e}}, \citenamefont {Mottershead}, \citenamefont {Elfimov},
  \citenamefont {Peets}, \citenamefont {Liang}, \citenamefont {Bonn},
  \citenamefont {Hardy}, \citenamefont {Chiuzbaian}, \citenamefont {Falub},
  \citenamefont {Shi}, \citenamefont {Patthey},\ and\ \citenamefont
  {Damascelli}}]{PlatePRL05}%
  \BibitemOpen
  \bibfield  {author} {\bibinfo {author} {\bibfnamefont {M.}~\bibnamefont
  {Plat{\'e}}}, \bibinfo {author} {\bibfnamefont {J.~D.~F.}\ \bibnamefont
  {Mottershead}}, \bibinfo {author} {\bibfnamefont {I.~S.}\ \bibnamefont
  {Elfimov}}, \bibinfo {author} {\bibfnamefont {D.~C.}\ \bibnamefont {Peets}},
  \bibinfo {author} {\bibfnamefont {R.}~\bibnamefont {Liang}}, \bibinfo
  {author} {\bibfnamefont {D.~A.}\ \bibnamefont {Bonn}}, \bibinfo {author}
  {\bibfnamefont {W.~N.}\ \bibnamefont {Hardy}}, \bibinfo {author}
  {\bibfnamefont {S.}~\bibnamefont {Chiuzbaian}}, \bibinfo {author}
  {\bibfnamefont {M.}~\bibnamefont {Falub}}, \bibinfo {author} {\bibfnamefont
  {M.}~\bibnamefont {Shi}}, \bibinfo {author} {\bibfnamefont {L.}~\bibnamefont
  {Patthey}}, \ and\ \bibinfo {author} {\bibfnamefont {A.}~\bibnamefont
  {Damascelli}},\ }\href {\doibase 10.1103/PhysRevLett.95.077001} {\bibfield
  {journal} {\bibinfo  {journal} {Phys. Rev. Lett.}\ }\textbf {\bibinfo
  {volume} {95}},\ \bibinfo {pages} {077001} (\bibinfo {year}
  {2005})}\BibitemShut {NoStop}%
\bibitem [{\citenamefont {Peets}\ \emph {et~al.}(2007)\citenamefont {Peets},
  \citenamefont {Mottershead}, \citenamefont {Wu}, \citenamefont {Elfimov},
  \citenamefont {Liang}, \citenamefont {Hardy}, \citenamefont {Bonn},
  \citenamefont {Raudsepp}, \citenamefont {Ingle},\ and\ \citenamefont
  {Damascelli}}]{PeetsNJP07}%
  \BibitemOpen
  \bibfield  {author} {\bibinfo {author} {\bibfnamefont {D.~C.}\ \bibnamefont
  {Peets}}, \bibinfo {author} {\bibfnamefont {J.~D.~F.}\ \bibnamefont
  {Mottershead}}, \bibinfo {author} {\bibfnamefont {B.}~\bibnamefont {Wu}},
  \bibinfo {author} {\bibfnamefont {I.~S.}\ \bibnamefont {Elfimov}}, \bibinfo
  {author} {\bibfnamefont {R.}~\bibnamefont {Liang}}, \bibinfo {author}
  {\bibfnamefont {W.~N.}\ \bibnamefont {Hardy}}, \bibinfo {author}
  {\bibfnamefont {D.~A.}\ \bibnamefont {Bonn}}, \bibinfo {author}
  {\bibfnamefont {M.}~\bibnamefont {Raudsepp}}, \bibinfo {author}
  {\bibfnamefont {N.~J.~C.}\ \bibnamefont {Ingle}}, \ and\ \bibinfo {author}
  {\bibfnamefont {A.}~\bibnamefont {Damascelli}},\ }\href
  {http://stacks.iop.org/1367-2630/9/i=2/a=028} {\bibfield  {journal} {\bibinfo
   {journal} {New Journal of Physics}\ }\textbf {\bibinfo {volume} {9}},\
  \bibinfo {pages} {28} (\bibinfo {year} {2007})}\BibitemShut {NoStop}%
\bibitem [{\citenamefont {Palczewski}\ \emph {et~al.}(2008)\citenamefont
  {Palczewski}, \citenamefont {Kondo}, \citenamefont {Khasanov}, \citenamefont
  {Kolesnikov}, \citenamefont {Timonina}, \citenamefont {Rotenberg},
  \citenamefont {Ohta}, \citenamefont {Bendounan}, \citenamefont {Sassa},
  \citenamefont {Fedorov}, \citenamefont {Pailh{\'e}s}, \citenamefont
  {Santander-Syro}, \citenamefont {Chang}, \citenamefont {Shi}, \citenamefont
  {Mesot}, \citenamefont {Fretwell},\ and\ \citenamefont
  {Kaminski}}]{PalczewskiPRB08}%
  \BibitemOpen
  \bibfield  {author} {\bibinfo {author} {\bibfnamefont {A.~D.}\ \bibnamefont
  {Palczewski}}, \bibinfo {author} {\bibfnamefont {T.}~\bibnamefont {Kondo}},
  \bibinfo {author} {\bibfnamefont {R.}~\bibnamefont {Khasanov}}, \bibinfo
  {author} {\bibfnamefont {N.~N.}\ \bibnamefont {Kolesnikov}}, \bibinfo
  {author} {\bibfnamefont {A.~V.}\ \bibnamefont {Timonina}}, \bibinfo {author}
  {\bibfnamefont {E.}~\bibnamefont {Rotenberg}}, \bibinfo {author}
  {\bibfnamefont {T.}~\bibnamefont {Ohta}}, \bibinfo {author} {\bibfnamefont
  {A.}~\bibnamefont {Bendounan}}, \bibinfo {author} {\bibfnamefont
  {Y.}~\bibnamefont {Sassa}}, \bibinfo {author} {\bibfnamefont
  {A.}~\bibnamefont {Fedorov}}, \bibinfo {author} {\bibfnamefont
  {S.}~\bibnamefont {Pailh{\'e}s}}, \bibinfo {author} {\bibfnamefont {A.~F.}\
  \bibnamefont {Santander-Syro}}, \bibinfo {author} {\bibfnamefont
  {J.}~\bibnamefont {Chang}}, \bibinfo {author} {\bibfnamefont
  {M.}~\bibnamefont {Shi}}, \bibinfo {author} {\bibfnamefont {J.}~\bibnamefont
  {Mesot}}, \bibinfo {author} {\bibfnamefont {H.~M.}\ \bibnamefont {Fretwell}},
  \ and\ \bibinfo {author} {\bibfnamefont {A.}~\bibnamefont {Kaminski}},\
  }\href {\doibase 10.1103/PhysRevB.78.054523} {\bibfield  {journal} {\bibinfo
  {journal} {Phys. Rev. B}\ }\textbf {\bibinfo {volume} {78}},\ \bibinfo
  {pages} {054523} (\bibinfo {year} {2008})}\BibitemShut {NoStop}%
\bibitem [{\citenamefont {Sakakibara}\ \emph {et~al.}(2012)\citenamefont
  {Sakakibara}, \citenamefont {Usui}, \citenamefont {Kuroki}, \citenamefont
  {Arita},\ and\ \citenamefont {Aoki}}]{SakakibaraPRB12}%
  \BibitemOpen
  \bibfield  {author} {\bibinfo {author} {\bibfnamefont {H.}~\bibnamefont
  {Sakakibara}}, \bibinfo {author} {\bibfnamefont {H.}~\bibnamefont {Usui}},
  \bibinfo {author} {\bibfnamefont {K.}~\bibnamefont {Kuroki}}, \bibinfo
  {author} {\bibfnamefont {R.}~\bibnamefont {Arita}}, \ and\ \bibinfo {author}
  {\bibfnamefont {H.}~\bibnamefont {Aoki}},\ }\href {\doibase
  10.1103/PhysRevB.85.064501} {\bibfield  {journal} {\bibinfo  {journal} {Phys.
  Rev. B}\ }\textbf {\bibinfo {volume} {85}},\ \bibinfo {pages} {064501}
  (\bibinfo {year} {2012})}\BibitemShut {NoStop}%
\bibitem [{\citenamefont {Wakimoto}\ \emph {et~al.}(2015)\citenamefont
  {Wakimoto}, \citenamefont {Ishii}, \citenamefont {Kimura}, \citenamefont
  {Fujita}, \citenamefont {Dellea}, \citenamefont {Kummer}, \citenamefont
  {Braicovich}, \citenamefont {Ghiringhelli}, \citenamefont {Debeer-Schmitt},\
  and\ \citenamefont {Granroth}}]{WakimotoPRB2015}%
  \BibitemOpen
  \bibfield  {author} {\bibinfo {author} {\bibfnamefont {S.}~\bibnamefont
  {Wakimoto}}, \bibinfo {author} {\bibfnamefont {K.}~\bibnamefont {Ishii}},
  \bibinfo {author} {\bibfnamefont {H.}~\bibnamefont {Kimura}}, \bibinfo
  {author} {\bibfnamefont {M.}~\bibnamefont {Fujita}}, \bibinfo {author}
  {\bibfnamefont {G.}~\bibnamefont {Dellea}}, \bibinfo {author} {\bibfnamefont
  {K.}~\bibnamefont {Kummer}}, \bibinfo {author} {\bibfnamefont
  {L.}~\bibnamefont {Braicovich}}, \bibinfo {author} {\bibfnamefont
  {G.}~\bibnamefont {Ghiringhelli}}, \bibinfo {author} {\bibfnamefont {L.~M.}\
  \bibnamefont {Debeer-Schmitt}}, \ and\ \bibinfo {author} {\bibfnamefont
  {G.~E.}\ \bibnamefont {Granroth}},\ }\href {\doibase
  10.1103/PhysRevB.91.184513} {\bibfield  {journal} {\bibinfo  {journal} {Phys.
  Rev. B}\ }\textbf {\bibinfo {volume} {91}},\ \bibinfo {pages} {184513}
  (\bibinfo {year} {2015})}\BibitemShut {NoStop}%
\bibitem [{\citenamefont {Meyers}\ \emph {et~al.}(2017)\citenamefont {Meyers},
  \citenamefont {Miao}, \citenamefont {Walters}, \citenamefont {Bisogni},
  \citenamefont {Springell}, \citenamefont {d'Astuto}, \citenamefont {Dantz},
  \citenamefont {Pelliciari}, \citenamefont {Huang}, \citenamefont {Okamoto},
  \citenamefont {Huang}, \citenamefont {Hill}, \citenamefont {He},
  \citenamefont {{Bo\ifmmode \check{z}\else \v{z}\fi{}ovi\ifmmode
  \acute{c}\else {\'c}\fi{}}}, \citenamefont {Schmitt},\ and\ \citenamefont
  {Dean}}]{Meyers16}%
  \BibitemOpen
  \bibfield  {author} {\bibinfo {author} {\bibfnamefont {D.}~\bibnamefont
  {Meyers}}, \bibinfo {author} {\bibfnamefont {H.}~\bibnamefont {Miao}},
  \bibinfo {author} {\bibfnamefont {A.~C.}\ \bibnamefont {Walters}}, \bibinfo
  {author} {\bibfnamefont {V.}~\bibnamefont {Bisogni}}, \bibinfo {author}
  {\bibfnamefont {R.~S.}\ \bibnamefont {Springell}}, \bibinfo {author}
  {\bibfnamefont {M.}~\bibnamefont {d'Astuto}}, \bibinfo {author}
  {\bibfnamefont {M.}~\bibnamefont {Dantz}}, \bibinfo {author} {\bibfnamefont
  {J.}~\bibnamefont {Pelliciari}}, \bibinfo {author} {\bibfnamefont {H.~Y.}\
  \bibnamefont {Huang}}, \bibinfo {author} {\bibfnamefont {J.}~\bibnamefont
  {Okamoto}}, \bibinfo {author} {\bibfnamefont {D.~J.}\ \bibnamefont {Huang}},
  \bibinfo {author} {\bibfnamefont {J.~P.}\ \bibnamefont {Hill}}, \bibinfo
  {author} {\bibfnamefont {X.}~\bibnamefont {He}}, \bibinfo {author}
  {\bibfnamefont {I.}~\bibnamefont {{Bo\ifmmode \check{z}\else
  \v{z}\fi{}ovi\ifmmode \acute{c}\else {\'c}\fi{}}}}, \bibinfo {author}
  {\bibfnamefont {T.}~\bibnamefont {Schmitt}}, \ and\ \bibinfo {author}
  {\bibfnamefont {M.~P.~M.}\ \bibnamefont {Dean}},\ }\href {\doibase
  10.1103/PhysRevB.95.075139} {\bibfield  {journal} {\bibinfo  {journal} {Phys.
  Rev. B}\ }\textbf {\bibinfo {volume} {95}},\ \bibinfo {pages} {075139}
  (\bibinfo {year} {2017})}\BibitemShut {NoStop}%
\bibitem [{\citenamefont {Locquet}\ \emph {et~al.}(1998)\citenamefont
  {Locquet}, \citenamefont {Perret}, \citenamefont {Fompeyrine}, \citenamefont
  {Machler}, \citenamefont {Seo},\ and\ \citenamefont {{Van
  Tendeloo}}}]{LocquetNat98}%
  \BibitemOpen
  \bibfield  {author} {\bibinfo {author} {\bibfnamefont {J.~P.}\ \bibnamefont
  {Locquet}}, \bibinfo {author} {\bibfnamefont {J.}~\bibnamefont {Perret}},
  \bibinfo {author} {\bibfnamefont {J.}~\bibnamefont {Fompeyrine}}, \bibinfo
  {author} {\bibfnamefont {E.}~\bibnamefont {Machler}}, \bibinfo {author}
  {\bibfnamefont {J.~W.}\ \bibnamefont {Seo}}, \ and\ \bibinfo {author}
  {\bibfnamefont {G.}~\bibnamefont {{Van Tendeloo}}},\ }\href
  {http://dx.doi.org/10.1038/28810} {\bibfield  {journal} {\bibinfo  {journal}
  {Nature}\ }\textbf {\bibinfo {volume} {394}},\ \bibinfo {pages} {453}
  (\bibinfo {year} {1998})}\BibitemShut {NoStop}%
\bibitem [{\citenamefont {Abrecht}\ \emph {et~al.}(2003)\citenamefont
  {Abrecht}, \citenamefont {Ariosa}, \citenamefont {Cloetta}, \citenamefont
  {Mitrovic}, \citenamefont {Onellion}, \citenamefont {Xi}, \citenamefont
  {Margaritondo},\ and\ \citenamefont {Pavuna}}]{AbrechtPRL03}%
  \BibitemOpen
  \bibfield  {author} {\bibinfo {author} {\bibfnamefont {M.}~\bibnamefont
  {Abrecht}}, \bibinfo {author} {\bibfnamefont {D.}~\bibnamefont {Ariosa}},
  \bibinfo {author} {\bibfnamefont {D.}~\bibnamefont {Cloetta}}, \bibinfo
  {author} {\bibfnamefont {S.}~\bibnamefont {Mitrovic}}, \bibinfo {author}
  {\bibfnamefont {M.}~\bibnamefont {Onellion}}, \bibinfo {author}
  {\bibfnamefont {X.~X.}\ \bibnamefont {Xi}}, \bibinfo {author} {\bibfnamefont
  {G.}~\bibnamefont {Margaritondo}}, \ and\ \bibinfo {author} {\bibfnamefont
  {D.}~\bibnamefont {Pavuna}},\ }\href {\doibase 10.1103/PhysRevLett.91.057002}
  {\bibfield  {journal} {\bibinfo  {journal} {Phys. Rev. Lett.}\ }\textbf
  {\bibinfo {volume} {91}},\ \bibinfo {pages} {057002} (\bibinfo {year}
  {2003})}\BibitemShut {NoStop}%
\bibitem [{\citenamefont {Radaelli}\ \emph {et~al.}(1994)\citenamefont
  {Radaelli}, \citenamefont {Hinks}, \citenamefont {Mitchell}, \citenamefont
  {Hunter}, \citenamefont {Wagner}, \citenamefont {Dabrowski}, \citenamefont
  {Vandervoort}, \citenamefont {Viswanathan},\ and\ \citenamefont
  {Jorgensen}}]{RadaelliPRB94}%
  \BibitemOpen
  \bibfield  {author} {\bibinfo {author} {\bibfnamefont {P.~G.}\ \bibnamefont
  {Radaelli}}, \bibinfo {author} {\bibfnamefont {D.~G.}\ \bibnamefont {Hinks}},
  \bibinfo {author} {\bibfnamefont {A.~W.}\ \bibnamefont {Mitchell}}, \bibinfo
  {author} {\bibfnamefont {B.~A.}\ \bibnamefont {Hunter}}, \bibinfo {author}
  {\bibfnamefont {J.~L.}\ \bibnamefont {Wagner}}, \bibinfo {author}
  {\bibfnamefont {B.}~\bibnamefont {Dabrowski}}, \bibinfo {author}
  {\bibfnamefont {K.~G.}\ \bibnamefont {Vandervoort}}, \bibinfo {author}
  {\bibfnamefont {H.~K.}\ \bibnamefont {Viswanathan}}, \ and\ \bibinfo {author}
  {\bibfnamefont {J.~D.}\ \bibnamefont {Jorgensen}},\ }\href {\doibase
  10.1103/PhysRevB.49.4163} {\bibfield  {journal} {\bibinfo  {journal} {Phys.
  Rev. B}\ }\textbf {\bibinfo {volume} {49}},\ \bibinfo {pages} {4163}
  (\bibinfo {year} {1994})}\BibitemShut {NoStop}%
\bibitem [{\citenamefont {Blaha}\ \emph {et~al.}(2001)\citenamefont {Blaha},
  \citenamefont {Schwarz}, \citenamefont {Madsen}, \citenamefont {Kvasnicka},\
  and\ \citenamefont {Luitz}}]{Blaha2001}%
  \BibitemOpen
  \bibfield  {author} {\bibinfo {author} {\bibfnamefont {P.}~\bibnamefont
  {Blaha}}, \bibinfo {author} {\bibfnamefont {K.}~\bibnamefont {Schwarz}},
  \bibinfo {author} {\bibfnamefont {G.}~\bibnamefont {Madsen}}, \bibinfo
  {author} {\bibfnamefont {D.}~\bibnamefont {Kvasnicka}}, \ and\ \bibinfo
  {author} {\bibfnamefont {J.}~\bibnamefont {Luitz}},\ }\href@noop {}
  {\bibfield  {journal} {\bibinfo  {journal} {Wien2k: An Augmented Plane Wave +
  Local Orbitals Program for Calculating Crystal Properties (Vienna University
  of Technology, Vienna, 2001)}\ } (\bibinfo {year} {2001})}\BibitemShut
  {NoStop}%
\bibitem [{\citenamefont {Perdew}\ \emph {et~al.}(1996)\citenamefont {Perdew},
  \citenamefont {Burke},\ and\ \citenamefont {Ernzerhof}}]{PerdewPRL1996}%
  \BibitemOpen
  \bibfield  {author} {\bibinfo {author} {\bibfnamefont {J.~P.}\ \bibnamefont
  {Perdew}}, \bibinfo {author} {\bibfnamefont {K.}~\bibnamefont {Burke}}, \
  and\ \bibinfo {author} {\bibfnamefont {M.}~\bibnamefont {Ernzerhof}},\ }\href
  {\doibase 10.1103/PhysRevLett.77.3865} {\bibfield  {journal} {\bibinfo
  {journal} {Phys. Rev. Lett.}\ }\textbf {\bibinfo {volume} {77}},\ \bibinfo
  {pages} {3865} (\bibinfo {year} {1996})}\BibitemShut {NoStop}%
\end{thebibliography}%

\end{document}